\documentclass[twocolumn]{aastex62}

\usepackage[utf8]{inputenc}
\usepackage{hyperref}
\usepackage{silence}
\usepackage{graphicx}
\usepackage[caption=false]{subfig}

\usepackage{float}
\usepackage{color,soul}

\usepackage{wrapfig}
\usepackage{bm}
\usepackage{amsmath}

\usepackage{fancyhdr}
\usepackage{wasysym}
\usepackage{natbib}

\begin{document}

\title{Environmental Effect on the Interstellar Medium in Galaxies across the Cosmic Web at $z=0.73$}

\author{S. K. Betti}
\affiliation{Department of Astronomy, University of Massachusetts, Amherst, MA 01003, USA}

\author{Alexandra Pope}
\affiliation{Department of Astronomy, University of Massachusetts, Amherst, MA 01003, USA}

\author{N. Scoville} 
\affiliation{Cahill Center for Astrophysics, California Institute of Technology, 1216 East California Boulevard, Pasadena, CA 91125, USA}

\author{Min S. Yun}
\affiliation{Department of Astronomy, University of Massachusetts, Amherst, MA 01003, USA}

\author{H. Aussel}
\affiliation{AIM Unit$\acute{e}$ Mixte de Recherche CEA CNRS, Universit$\acute{e}$ Paris VII UMR n158, Paris, France}

\author{J. Kartaltepe}
\affiliation{School of Physics and Astronomy, Rochester Institute of Technology, 84 Lomb Memorial Drive, Rochester, NY 14623, USA} 

\author{K. Sheth}
\affiliation{NASA Headquarters, 300 E Street SW, Washington DC 20546, USA}

\correspondingauthor{S. K. Betti}
\email{sbetti@umass.edu}

\singlespace 
\begin{abstract}
We present new ALMA dust continuum observations of 101 $\log(\mathrm{M}_* / \mathrm{M}_\odot) > 9.5$  galaxies in the COSMOS field to study the effect of environment on the interstellar medium at $z\sim0.7$.  At this redshift, our targets span a wide range of environments allowing for a diverse sample of galaxies with densities, $\Sigma = 0.16-10.5 \ \mathrm{Mpc}^{-2}$  (per $\Delta z = 0.024$).  
Using the ALMA observations, we calculate the total ISM mass ($\mathrm{M}_{\mathrm{ISM}}$) and look for depletion as a function of galaxy density in order to understand the quenching or triggering of star formation in galaxies in different environments.  $\mathrm{M}_{\mathrm{ISM}}$ is found to have a small dependence on environment, while the depletion timescale remains constant ($\sim200$ Myrs) across all environments.  We find elevated $\mathrm{M}_\mathrm{ISM}$ values at intermediate densities and lower values at high densities compared to low (field) densities.  Our observed evolution in gas fraction with density in this single redshift slice is equivalent to the observed evolution with cosmic time over $2-3$ Gyr.  To explain the change in gas mass fraction seen in galaxies in intermediate and high densities, these results suggest environmental processes such as mergers and ram pressure stripping are likely playing a role in dense filamentary-cluster environments. 
\end{abstract}
\doublespace


\section{Introduction}
It is well known that environment plays a role in influencing physical processes of galaxies \citep[e.g.][]{Dressler1980, Kauffmann2004, Peng2010}.  
In the local Universe, galaxies in dense environments are generally early-type, red-sequence massive galaxies \citep{Peng2010} in a state of passive evolution with little star formation.  However, at higher redshift ($z\gtrsim1$), ongoing star formation has been found in galaxies in mid to dense environments \citep{Alberts2014, Tran2010, Brodwin2013} with increasing densities having little effect on the average star formation rate (SFR) of galaxies \citep{Elbaz2007, Cooper2008, Scoville2013}.

At all redshifts and environments, molecular gas in the interstellar medium (ISM) of galaxies is fuel for star formation.  However, the cause of the change from active star forming galaxies to passive galaxies after the peak epoch of star formation, especially in dense environments, is still unknown.  Some recent work has suggested the cause of the decline in star formation after $z\sim2$ \citep{Madau2014} to be a decrease in star formation efficiency \citep[SFE;][]{Santini2014} though others have found a weaker evolution of SFE with star formation \citep{Saintonge2013, Dessauges-Zavadsky2015, Bethermin2015, Scoville2017}.  Along with SFE, a decrease in gas accretion and depletion time has been found to depend on cosmic time \citep{Bouche2010, Lilly2013, Bethermin2013, Behroozi2013,Genzel2015, Scoville2017, Tacconi2018}.

\begin{figure}[tb]
\centering
\begin{minipage}{.47\textwidth}
\includegraphics[width=\linewidth]{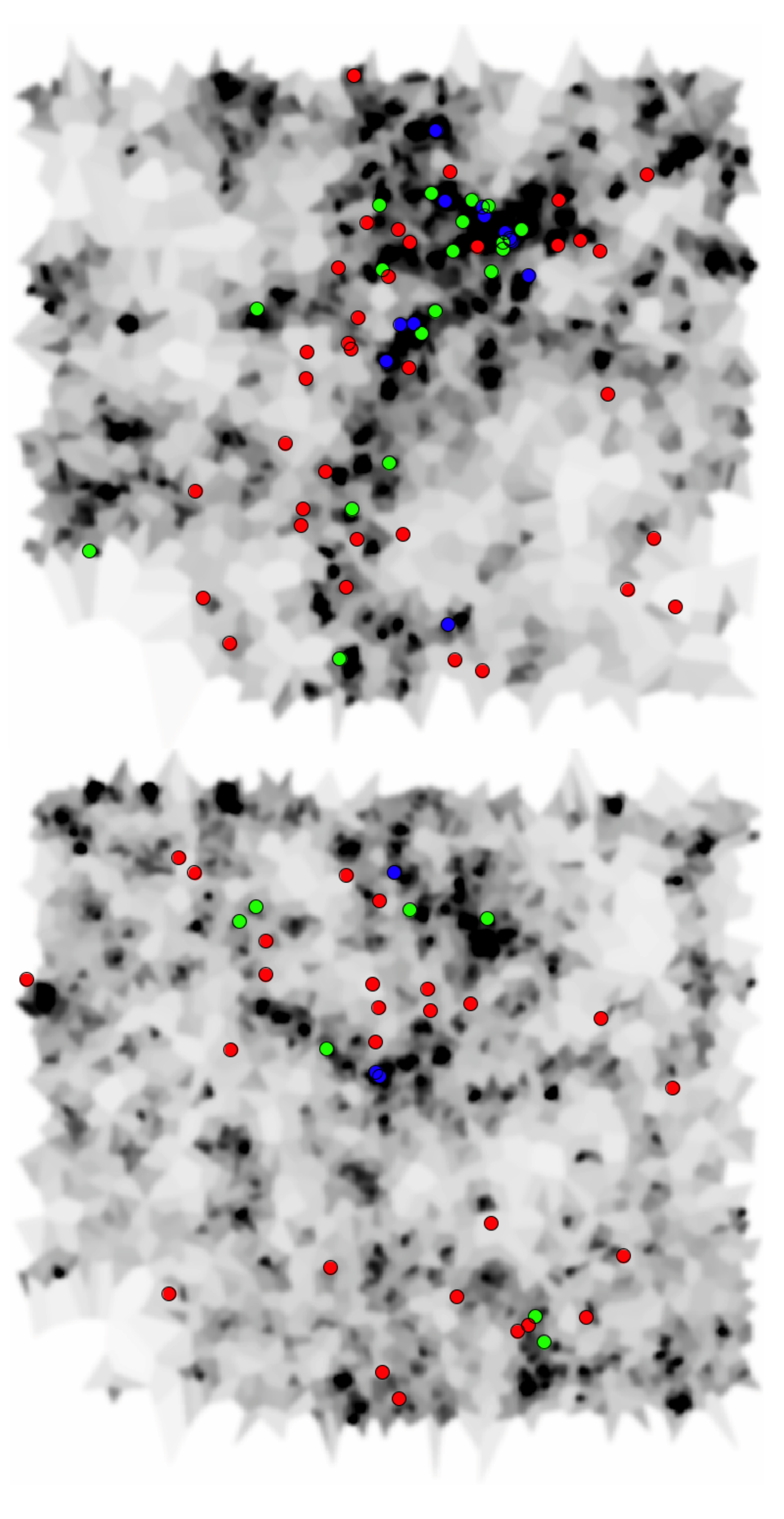}
\end{minipage}
\caption{Density maps from \citet{Scoville2013} for $z = 0.715 - 0.739$ (top) and $z = 0.739 - 0.764$ (bottom) (dark corresponds to higher density regions) with our ALMA galaxy sample overlaid.  The colored circles correspond to galaxies in local galaxy density ($\Sigma$ [Mpc$^{-2}$]) bins with red: $ \Sigma < 1.2 \ \mathrm{Mpc}^{-2}$, green: $1.2 \ \mathrm{Mpc}^{-2} < \Sigma < 2.6 \ \mathrm{Mpc}^{-2}$, blue: $\Sigma > 2.6\ \mathrm{Mpc}^{-2}$.  Images are $1.4^\circ \times 1.4^\circ$.}
\label{diffz}
\end{figure}

Local passive galaxies have been studied to see if the environment has an effect on molecular gas, and if this can explain the decrease in star formation.  Early studies suggest the environment does not affect molecular gas as it is gravitationally bound to the center of the galaxy \citep{Casoli1991, Boselli1997, Lavezzi1998}.  However, due to advancements in detecting and measuring gas content through both CO and dust continuum observations, studies have found that molecular gas is being stripped in cluster galaxies \citep{Corbelli2012, Fumagalli2013, Jablonka2013} which could affect the SFE \citep{Mok2017, LeeB2017, Ebeling2014, Koyama2017}.  The effect of stripping the molecular gas allows the galaxies to remain in a state of passive evolution, and could lead to the anti-correlation between star formation and galaxy density in the local Universe. 

For this anti-correlation between galaxy density and star formation to hold true at low redshift, some high redshift galaxies must be in the process of quenching their star formation as star forming galaxies have been found in all environments at $z \geq 0.7$ \citep[e.g.][]{Scoville2013, Alberts2014}.
Around this epoch ($z\sim0.7$), galaxies appear to be switching from having environmentally free star formation to being dependent on environment.  In order to understand what drives this environmental dependency, observations of the ISM are key to determining how star formation is cut off in dense environments.

In this paper, we study 101 galaxies in the COSMOS 2 deg$^2$ survey \citep{Scoville2007a} at $z \sim0.7$.  Using observations of the dust continuum from the Atacama Large Millimeter Array (ALMA), we calculate the total ISM mass ($M_\mathrm{ISM}$) and look for depletion as a function of galaxy density in order to determine how the environment affects the evolution of galaxies.  

In Section 2, we discuss the sample selection and a priori properties of the sample, and in Section 3 we present the new ALMA observations.  In Section 4, we calculate the flux and mass measurements and perform a stacking analysis.  Our results are presented in Section 5 which we then discuss in Section 6.  We conclude in Section 7 by summarizing the main results.  
We assume a flat $\Lambda$CDM cosmology with $\Omega_m = 0.307$, $\Lambda$ = 0.7 and $H_0 = 67.7$ km s$^{-1}$ Mpc$^{-1}$ \citep{Planck2015}.  A \citet{Chabrier2003} IMF is used when deriving SFRs and stellar masses.

 \begin{figure*}[tb]
\centering
\includegraphics[width=\linewidth]{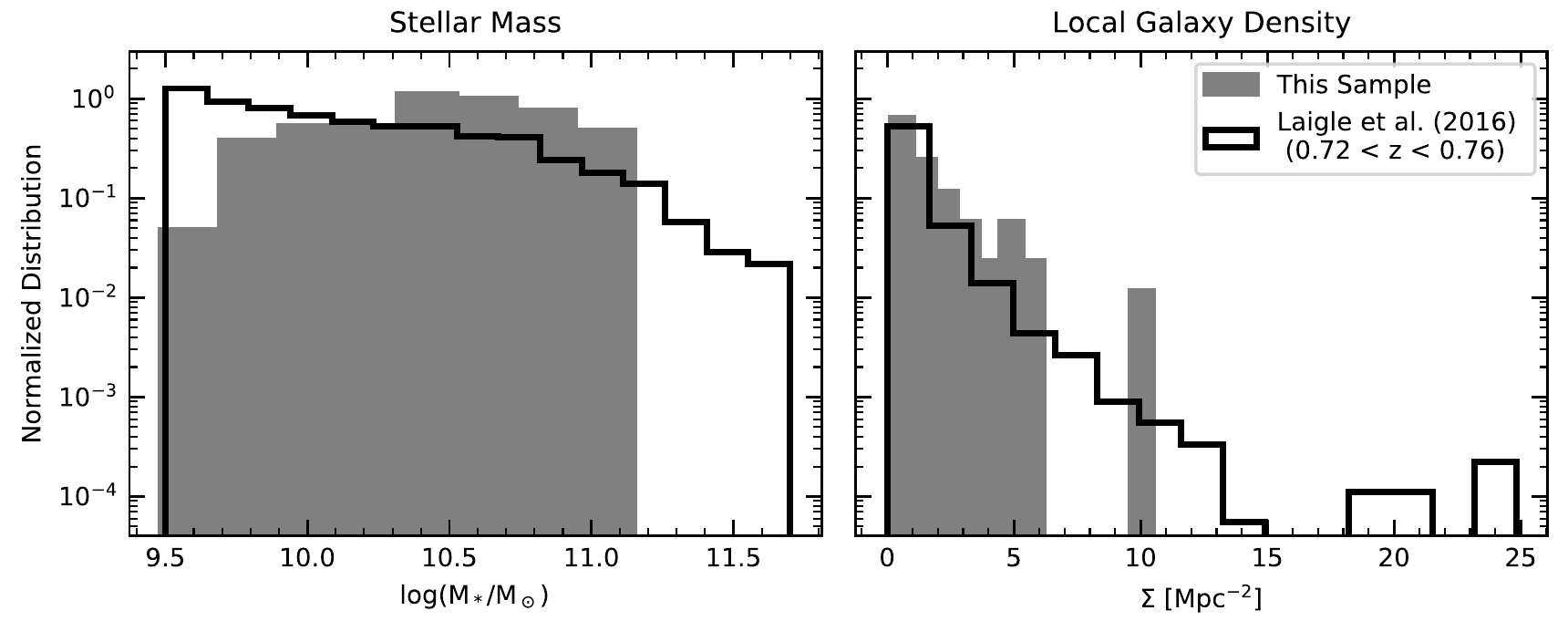}
\caption{Stellar mass ({\it left}) and local galaxy density ({\it right}) for our ALMA sample compared to the overall COSMOS photometric redshift sample at $z=0.72-0.76$ from \citet{Laigle2016}. }
\label{histograms}
\end{figure*}

\section{Sample}
Our sample of 101 galaxies is selected from the COSMOS 2 deg$^2$ survey \citep{Scoville2007a} which has multiwavelength coverage from 37 bands, including deep \textit{Herschel} (PACS and SPIRE) imaging from $100-500 \ \mu$m \citep{Oliver2010, Lutz2011}.  Accurate photometric redshifts in the COSMOS field have been derived from UV through near IR photometry from 34 bands, which is described in detail by \citet{Ilbert2013} and \citet{Laigle2016}.      

We started with all galaxies in the COSMOS field with 100 $\mu$m detections and spectroscopic redshifts obtained from the VLT-VIMOS zCOSMOS survey \citep{Lilly2007}.  
In order to sample a wide range of environments near a redshift range where local density starts to affect SFR, the spectroscopic redshift range $z = 0.72-0.76$ was selected as it shows a known large scale structure (LSS) in COSMOS \citep{Guzzo2007, Scoville2007b}.  The galaxies in this redshift range were then chosen if they have $S_{100 \ \mu\mathrm{m}} > 5$ mJy (equivalent to L$_\mathrm{IR} > 1.5 \times 10^{11}$ L$_{\odot}$ or SFR $> 20$ M/yr at $z=0.7$) in order to ensure they are star forming.  This sample selection criteria was used as 100 $\mu$m is the most sensitive band in the IR which also traces the total IR luminosity in the COSMOS field at $z\sim 0.7$ \citep{Elbaz2011}.  At 60 $\mu$m rest frame, we are probing the warmer dust; however, studies have shown that FIR wavelengths can be unbiased to all ULIRGs regardless of temperature \citep[see][]{Symeonidis2011}. Given that almost all of our sources (98/101) are also detected at 250 $\mu$m (150 $\mu$m rest frame), we are confident that we are selecting typical IR luminous galaxies at this epoch.

The local galaxy densities were determined from projected 2D density maps published by \citet{Scoville2013}, which mapped the COSMOS field out to $z \sim 3$.  \citet{Scoville2013} found that the projected 2D densities are related to the true 3D densities as long as the slices in redshift ($\Delta z$) are thin enough that there is no superposition of galaxies on the LSS from neighboring redshift bins.  Using adaptive smoothing and Voronoi tessellation on 155,954 $K_s$-band selected galaxies at $z=0.15-3.0$ from Ultra-Vista with photometric redshifts from \citet{Ilbert2013}, \citet{Scoville2013} mapped the cosmic LSS and estimated environmental densities for 127 redshift slices.  Herein, we will refer to these projected 2D densities as local galaxy density ($\Sigma$) given in comoving Mpc$^{-2}$.  In these maps, 250 significant overdense structures were found from filamentary to circularly symmetric, including a notable overdense structure at $z\sim0.7$.  

\begin{figure*}[tb]
\centering
\includegraphics[width=\linewidth]{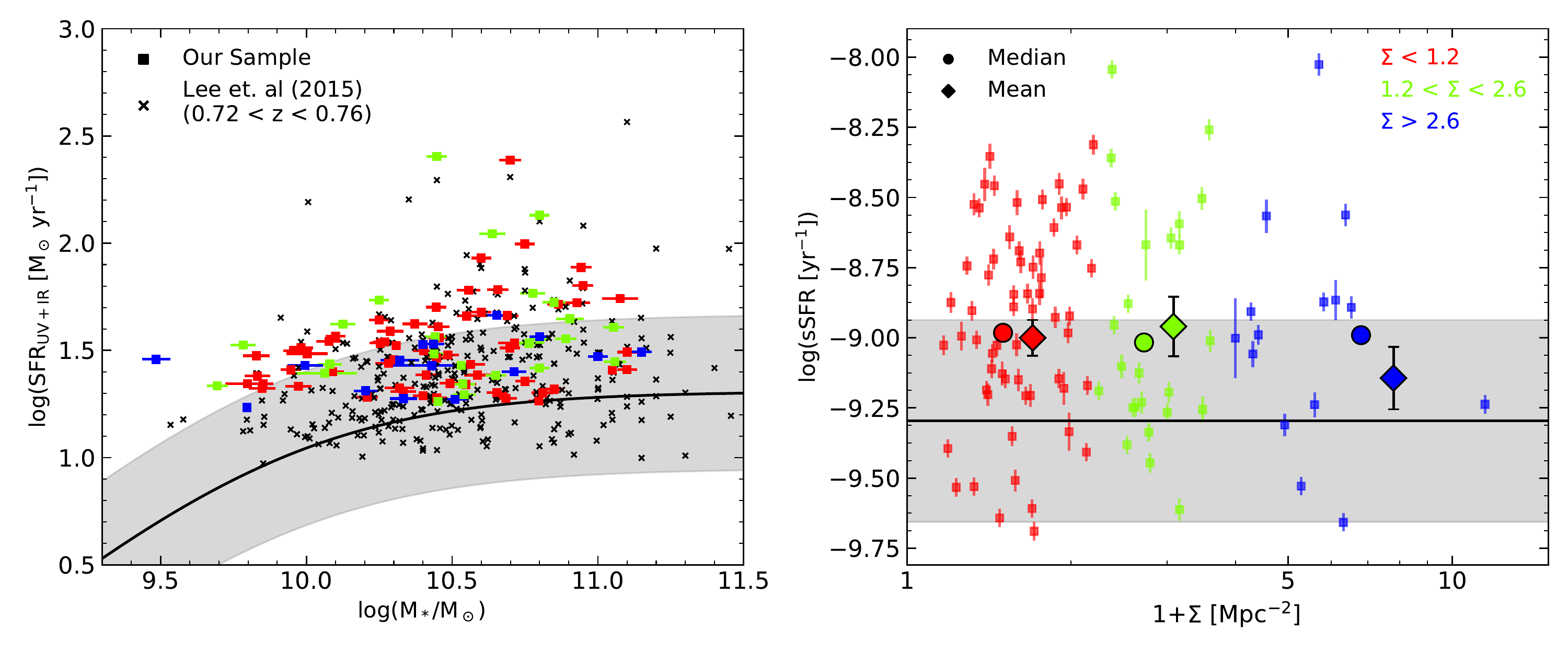}
\caption{Left: Star formation rate as a function of stellar mass for our sample (squares) and the overall sample from $z=0.72-0.76$ from the catalog by \citet{Lee2015} (x's).  The black curve is the MS line for $z = 0.7$ from \citet{Lee2015} with the associated error of $\sigma = 0.36$ dex.  Right: sSFR (SFR/M$_*$) as a function of density for our 101 galaxies (squares), the medians (circles) and means (diamonds) for each density bin.  The black line and grey band corresponds to the MS line from the {\it left} panel.  As the SFR and stellar mass was controlled for in selecting our sample, there is no significant dependence of sSFR with environment.}
\label{SFRandsSFR}
\end{figure*}

For our ALMA study, the redshift range $z = 0.72-0.76$ was chosen for the known LSS and therefore wide range in densities which allows us to probe a variety of environments.  In this redshift range, we found 101 galaxies with $\log(\mathrm{M}_*/\mathrm{M}_\odot) > 9.4$ (the mass completeness limit found by \citet{Laigle2016} is $\log(\mathrm{M}_*/\mathrm{M}_\odot) > 9.3$) and far-IR detections.  Over this redshift range, the projected 2D density slices have a thickness of $\Delta z = 0.024$; we show sample slices at $z = 0.715-0.739$ and $z=0.739-0.764$ along with our ALMA targets in Figure \ref{diffz} to highlight the broad range in projected 2D densities.      
We show the distribution of stellar mass and local galaxy density of these targets compared to the overall COSMOS sample from \citet{Laigle2016} for $z=0.72-0.76$ in Figure \ref{histograms}.  Stellar mass ($\log(\mathrm{M}_*/\mathrm{M}_\odot) > 9.4-11.0$) was controlled for across different densities (e.g.~Figure \ref{SFRandsSFR} {\it left}).

We used the IR photometry for our 101 sources from the PEP PACS catalogs \citep{Lutz2011}, which includes fluxes at 24 $\mu$m (Spitzer/MIPS) and 100 $\mu$m (\textit{Herschel}/PACS). In order to calculate the total IR luminosity (L$_{\mathrm{IR}}$, 8-1000 $\mu$m) for each galaxy, we fit the available IR photometry to the \citet{Kirkpatrick2015} spectral energy distribution (SED) templates which have been empirically derived for high redshift galaxies. We find that the AGN and composite galaxy templates are a poor fit to the data, especially when considering the 24/100 $\mu$m flux ratio. Using the SFG template appropriate for our $z\sim0.7$ targets, we derive L$_{\mathrm{IR}}$ values for all 101 galaxies.  

We compare our derived L$_{\mathrm{IR}}$ with L$_{\mathrm{IR}}$ values from the \citet{Lee2015} catalog who derived L$_{\mathrm{IR}}$ and L$_{\mathrm{UV}}$s for 4218 \textit{Herschel}-selected COSMOS sources in the redshift range $z = 0.02-3.54$.  \cite{Lee2013} determined IR luminosity by fitting a modified blackbody plus mid IR power law to full IR photometry following \citet{Casey2012}.  
Of the 101 selected galaxies, 80 had matches with the \citet{Lee2015} catalog.  The average percent difference in L$_\mathrm{IR}$ for the 80 matches was 15$\%$ (\citealt{Lee2015} lower by 15$\%$). This 15\% difference does not depend on density and the results of this paper are unchanged whether we use our derived L$_{\rm IR}$ values or the \citet{Lee2015} L$_{\rm IR}$ values.     

From L$_{\mathrm{IR}}$, we calculate SFR$_{\mathrm{IR}}$ using the relation from \citet{Arnouts2013}, who adopted the relation from \citet{Bell2005} after adjusting for a \citet{Chabrier2003} IMF,
\begin{equation}
\mathrm{SFR}_{\mathrm{IR}} = (8.6 \times 10^{-11}) \times L_{\mathrm{IR}}.
\end{equation}
The total SFR is derived from the sum of SFR$_{\mathrm{UV}}$ \citep[$L_\mathrm{UV} = \nu L_\nu(2300$ \AA)]{Lee2015} and SFR$_{\mathrm{IR}}$.  All 101 target galaxies have corresponding SFR$_\mathrm{UV}$ from \citet{Lee2015}.  The left panel of Figure \ref{SFRandsSFR} shows where the SFRs derived for our sample fall within the full sample from \citet{Lee2015} (adjusted for 15$\%$ difference in L$_\mathrm{IR}$).
We find that our sample falls slightly above the MS line calculated by \citet{Lee2015} for the redshift range $z = 0.63-0.78$ (the black curves in both panels in Figure \ref{SFRandsSFR}), and the sSFR (sSFR = SFR/M$_*$) remains relatively constant across all densities (right panel of Figure \ref{SFRandsSFR}).       

\section{Observations}
We measure the dust continuum for our sample with ALMA Cycle 3 observations (2015.1.00055.S; PI Pope).  The 101 galaxies which fit the criteria listed in the previous section were separated into two ALMA science goals: galaxies with $z = 0.72-0.741$ in the first science goal (SG1, 65 sources), and galaxies with $z = 0.741-0.76$ in the second science goal (SG2, 36 sources).  Both science goals were observed in Band 7 with SG1 observations taken between January $2 - 5$, 2016 ($\nu=345.7\,$GHz, bandwidth of 7.475 GHz) and SG2 observations taken between January 26 $-$ April 27, 2016 ($\nu=342.3\,$GHz, bandwidth of 7.425 GHz).  
On source integration time for both science goals was 3 minutes per galaxy.  
We used the delivered calibrated data for SG1, but had to manually recalibrate SG2 with the Common Astronomy Software Application \citep[CASA;][]{McMullin2007} due to issued flagged by the pipeline reduction. The data was then cleaned and imaged with CASA.  The average 1$\sigma$ rms sensitivity achieved is 0.15 mJy/beam with a beam size of $0.9\arcsec \times 0.5\arcsec$ for SG1 and 0.21 mJy/beam with a beam size of $1.0\arcsec \times 0.8\arcsec$ for SG2.  For each source, we made continuum maps and primary beam corrected continuum maps with the CASA task CLEAN using natural weighting and a threshold of 0.4 mJy ($2-3\sigma$).  The maps have a pixel scale of $0.12\arcsec$.

We repeated the imaging with the same weighting and threshold but with a Gaussian uv-taper for the higher resolution SG1 in order to match the beam size of SG2.  By lowering the resolution of SG1, we both ensure that any extended flux is not resolved out and that the two SG can be stacked without differing beam size affecting the integrated aperture flux measurements.  The average rms of the SG1 tapered images is 0.16 mJy/beam.  The uv-tapered SG1 maps are then used for both the individual and stacked measurements and analysis. RMS values for all 101 images are given in Table \ref{INFO}.    


\section{Analysis}     
\subsection{Flux Measurements}\label{FluxMeasurementSection}
\begin{figure}[!tb]
\centering
\includegraphics[width=\linewidth]{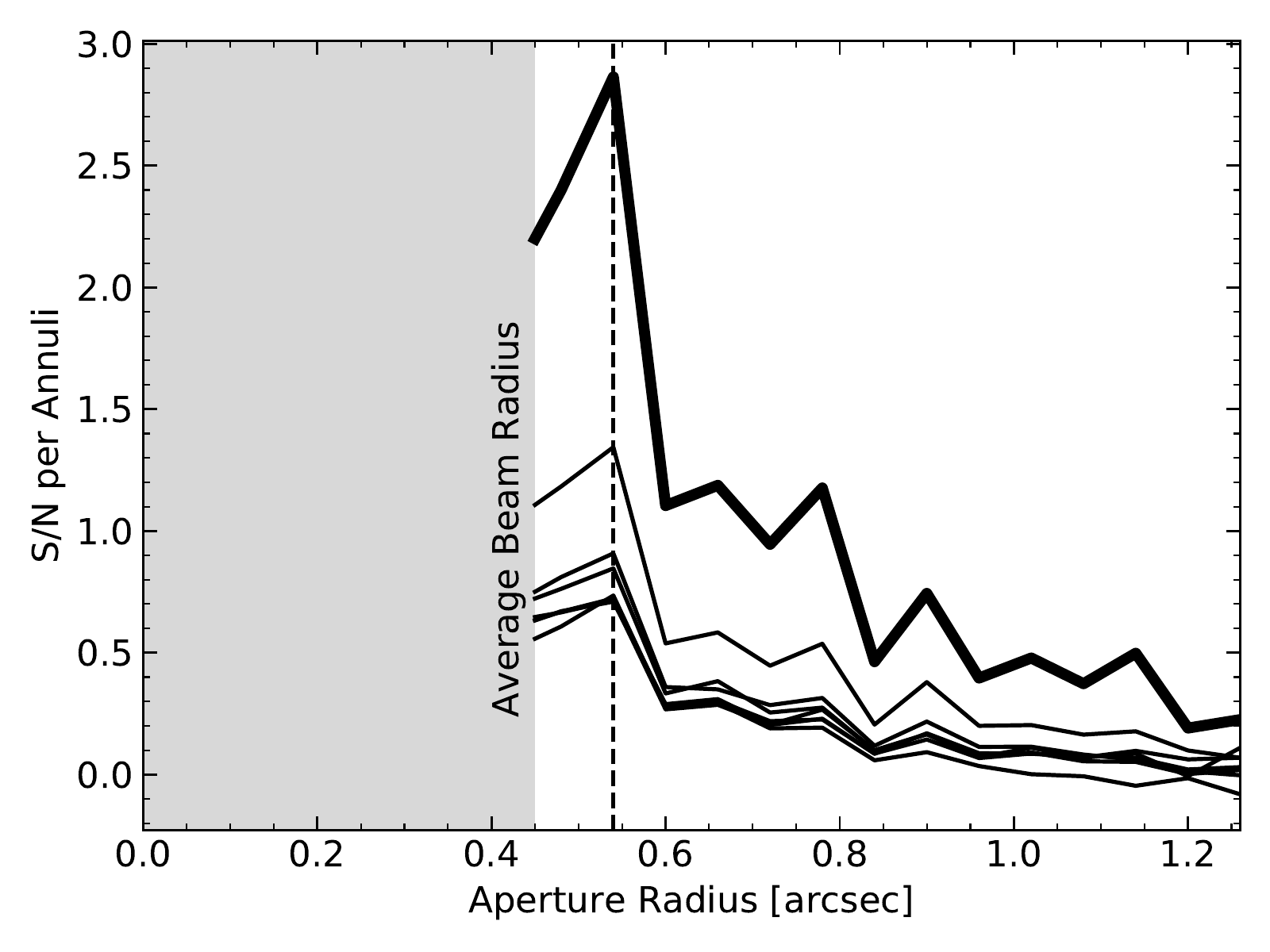}
 \caption{S/N per annuli as a function of aperture radius.  The thin lines indicate seven bright sources and the thick line is for all 101 sources mean stacked together. The grey band represents radii below the average radius of the beam. 
 In the stack and brightest sources, the optimal S/N per annuli is at a radius of $r = 0.54\arcsec$ (dotted vertical line), beyond which the S/N drops off rapidly.}
 \label{FluxvsRad-8brightest}
\end{figure}

We center an aperture at the center of the known optical position of each galaxy on the primary beam corrected maps to calculate the integrated aperture flux (S$_\mathrm{tot}$) and the highest single pixel peak flux (S$_\mathrm{pix}$).  The integrated flux captures extended flux beyond the beam, while peak pixel flux is best for unresolved emission.  
In order to optimize the S/N for S$_\mathrm{tot}$, we calculate the S/N per annuli for increasing aperture radii in order to determine the aperture radius which encloses the maximum signal and least noise.  The S/N per annuli will increase as more signal is enclosed relative to the noise; the S/N will reach a maximum at some radii beyond which it drops rapidly as noise dominates the annuli flux.  We only look at radii greater than the average beam as aperture radii below the size of the beam will exclude flux as it is spread across the beam.  Figure \ref{FluxvsRad-8brightest} shows the S/N per annuli for seven bright sources and the mean stack of all 101 galaxies (bold).  We look at bright sources in order to see the clear drop in S/N.  We find an optimal aperture with radius $0.54\arcsec$, which is used for all individual flux measurements.
The integrated flux signal measurement is found from the primary beam corrected map;
\begin{equation}
S_{\mathrm{tot}} = \frac{\sum_{i=0}^{N} S_{i}}{\mathrm{\# \ pixels/beam}},
\end{equation}
where $S_i$ is the flux in mJy/beam from each pixel within the aperture.  Dividing the summed flux in the aperture by $\#$ pixels/beam converts $S_{tot}$ from mJy/beam to mJy/pixel.

For our data, $\#$ pixels/beam is $\sim$ 66.14 pixels/beam.  The noise estimate for the integrated flux measurement, $\sigma_\mathrm{tot}$, are derived from the non primary beam corrected maps by taking the standard deviation of integrated flux measurements in 100 random apertures of the same size offset from the source.  

The peak flux signal measurement is found from the primary beam corrected map as the peak pixel in an aperture centered on the source.  The noise estimate, $\sigma_\mathrm{pix}$, is found by taking the average peak flux measurement of 100 apertures.  
Similar to \citet{Scoville2014}, a detection requires a $> 2\sigma$ integrated aperture flux measurement or, if the S/N$_{\mathrm{tot}} < 2$, we require a $3\sigma$ peak flux measurement.  All sources which have a $> 2\sigma$ integrated aperture flux also have a $> 3\sigma$ peak flux measurement.  

In SG2, two sources were located $\approx6\arcsec$ away from each other (source IDs: 32328 and 32520); however, as the beam size is $\sim1\arcsec$ there should not be any blending issues.  Due to their close location, these two sources were reimaged together in CASA using the same cleaning parameters described above in order to increase the S/N of each.  By imaging the sources together, the rms decreased from $\approx$0.214 mJy/beam to 0.152 mJy/beam and both sources were significantly detected in total aperture and peak pixel flux.    

Of the 101 galaxies, 68 are significantly detected in our ALMA band 7 data.  The 345 GHz fluxes range from $0.26-1.2$ mJy and the average flux is 0.45 mJy.  More details on the individual detections are given in Section \ref{5.1}.    
\subsection{ISM Mass}  
Though molecular gas at high redshift can be measured with CO observations \citep[see][]{Tacconi2010, Daddi2010, Genzel2010, Carilli2013},
estimating the ISM mass from CO lines is expensive and uncertain due to the poorly constrained conversion from CO to H$_{2}$.  In this study, we exploit the long wavelength Rayleigh-Jeans tail and use the dust continuum emission as a tracer of ISM mass \citep{Scoville2014, Eales2012, Magdis2013, Santini2014}.  This method was theorized by \citet{Hildebrand1983} who suggested that the dust mass of a galaxy could be estimated from submillimeter (submm) dust continuum emission.
Cold dust dominates the long wavelength Rayleigh-Jeans tail and it is assumed that this dust is in radiative equilibrium and is optically thin.  
As the dust is optically thin, the total dust content of the galaxy can be measured, and if the dust-to-gas ratio is assumed, $M_\mathrm{ISM}$ can be estimated.  

From local observations, both the dust emissivity per unit mass and the dust-to-gas ratio are constrained \citep[see][]{Draine2007,Galametz2011}.  In order to avoid a priori knowledge of the dust emissivity and dust-to-gas ratio, \citet{Scoville2014, Scoville2016} 
used local star-forming spirals, ultra-luminous infrared galaxies, and high-z submm galaxy samples, to empirically calibrate a ratio of the specific luminosity at a rest frame of 850 $\mu$m to the CO-derived (via $J=1\rightarrow 0$) $M_\mathrm{ISM}$ and found a single calibration constant 
\begin{equation}
\alpha_{850 \mu m} = \frac{L_{\nu 850\ \mu m}}{M_{ISM}} = 6.7 \times 10^{19} \ \mathrm{erg} \ \mathrm{s}^{-1} \ \mathrm{Hz}^{-1}  \ M_{\odot}^{-1}.
\end{equation}
Given the distance to the source, the flux density can be used to derive $M_\mathrm{ISM}$ of galaxies as 
\begin{equation} \label{MassISM}
\begin{split}
M_{\mathrm{ISM}} = &1.78 S_{\nu_{\mathrm{obs}}}[\mathrm{mJy}] (1+z)^{-4.8} \\
& \times \left(\frac{\nu_{850 \ \mu \mathrm{m}}}{\nu_{\mathrm{obs}}}\right)^{3.8} (d_L [\mathrm{Gpc}])^2 \\
& \times \left(\frac{6.7 \times 10^{19}}{\alpha_{850}} \right) \frac{\Gamma_{\mathrm{RJ}}}{\Gamma_0} 10^{10} M_{\odot} \\
& \mathrm{for }\ \lambda_\mathrm{rest} > 250 \ \mu \mathrm{m}, 
\end{split}
\end{equation}
where $\Gamma_{\mathrm{RJ}}$ is a correction factor for any Rayleigh-Jeans departures given by
\begin{equation}
\Gamma_{\mathrm{RJ}} (T_d, \nu_{\mathrm{obs}}, z) = \frac{h\nu_{\mathrm{obs}} (1+z) /kT_d}{e^{h\nu_{\mathrm{obs}} (1+z) / kT_d} - 1}.
\end{equation}
Following \citet{Scoville2014, Scoville2016}, we assume the temperature to be $T_d = 25$ K, which is observed to be a good estimate for high $z$ galaxies \citep[e.g.][]{Kirkpatrick2015}.  The rest wavelength is restricted to $\lambda_{\mathrm{rest}} > 250 \ \mu\mathrm{m}$ in order to constrain the dust to the Rayleigh-Jeans tail where emission is optically thin.  
For a complete derivation of the above equations, see \citet{Scoville2014, Scoville2016}.     

\subsection{Stacking Analysis} \label{4.3}
We stack our sample (both detections and non-detections) as a function of local galaxy density ($\Sigma$) in order to find the average submm flux density. 
The stacked flux densities can then be used to calculate how the mean ISM mass varies with environment.
We use the integrated aperture flux to measure the continuum in each stacked image.  In order to determine the size of the different stacked bins, we perform a sliding boxcar average of the integrated flux densities of all 101 galaxies as a function of their local galaxy density to determine the optimal subsamples to stack.  

\begin{figure}[tb]
\centering
\includegraphics[width=1\linewidth]{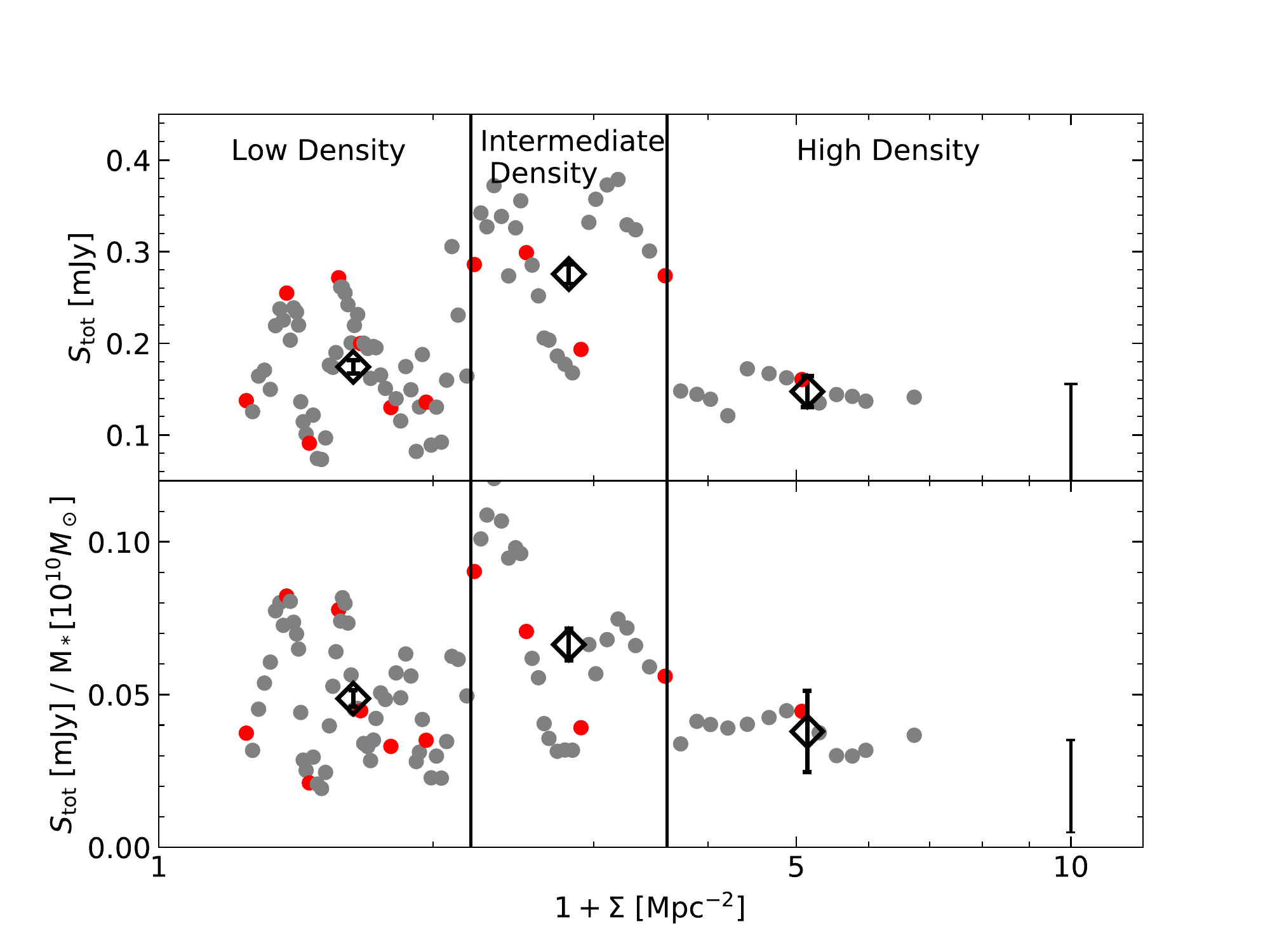}
\caption{Eight sample sliding boxcar averages for all 101 galaxies.  The bar indicates the error associated with each stack of 8 while the black diamond indicates the average in each density bin.  The red circles indicate independent measurements (every 8 points).  The bottom panel shows the flux per stellar mass for the same eight sample sliding boxcar average.  From both the flux and flux per M$_*$, our sample separates easily into 3 groups: $<1.2$, $1.2-2.6$, $>2.6$ galaxies Mpc$^{-2}$.}
\label{boxcarStacks}
\end{figure}  

\begin{figure*}[tb]
\hspace{-1.5cm}
\includegraphics[width=1.15\linewidth]{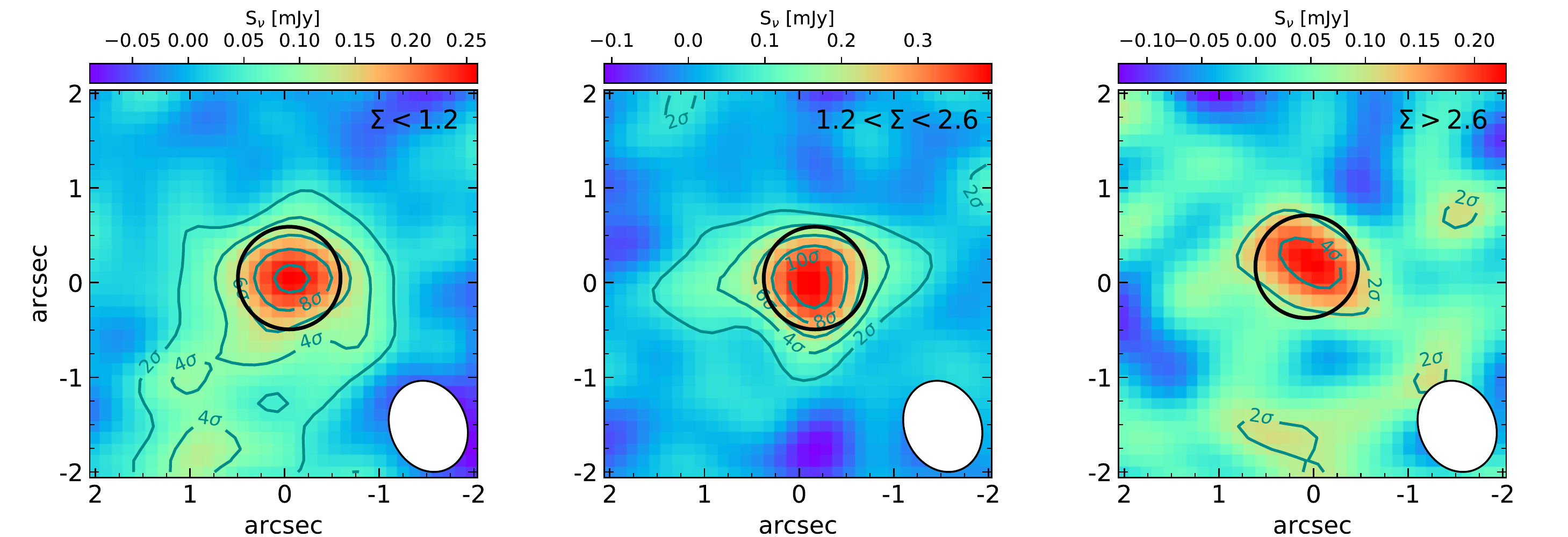}
\caption{Weighted mean stacked images for the three density bins.  The black circle is the aperture, $r = 0.54\arcsec$ used to determine $S_\mathrm{tot}$.  Contours show the SNR relative to the RMS in each stacked image.  The beam is shown in the lower right corner.} 
\label{stacks}
\end{figure*} 

\begin{figure}[tb]
\centering
\includegraphics[width=\linewidth]{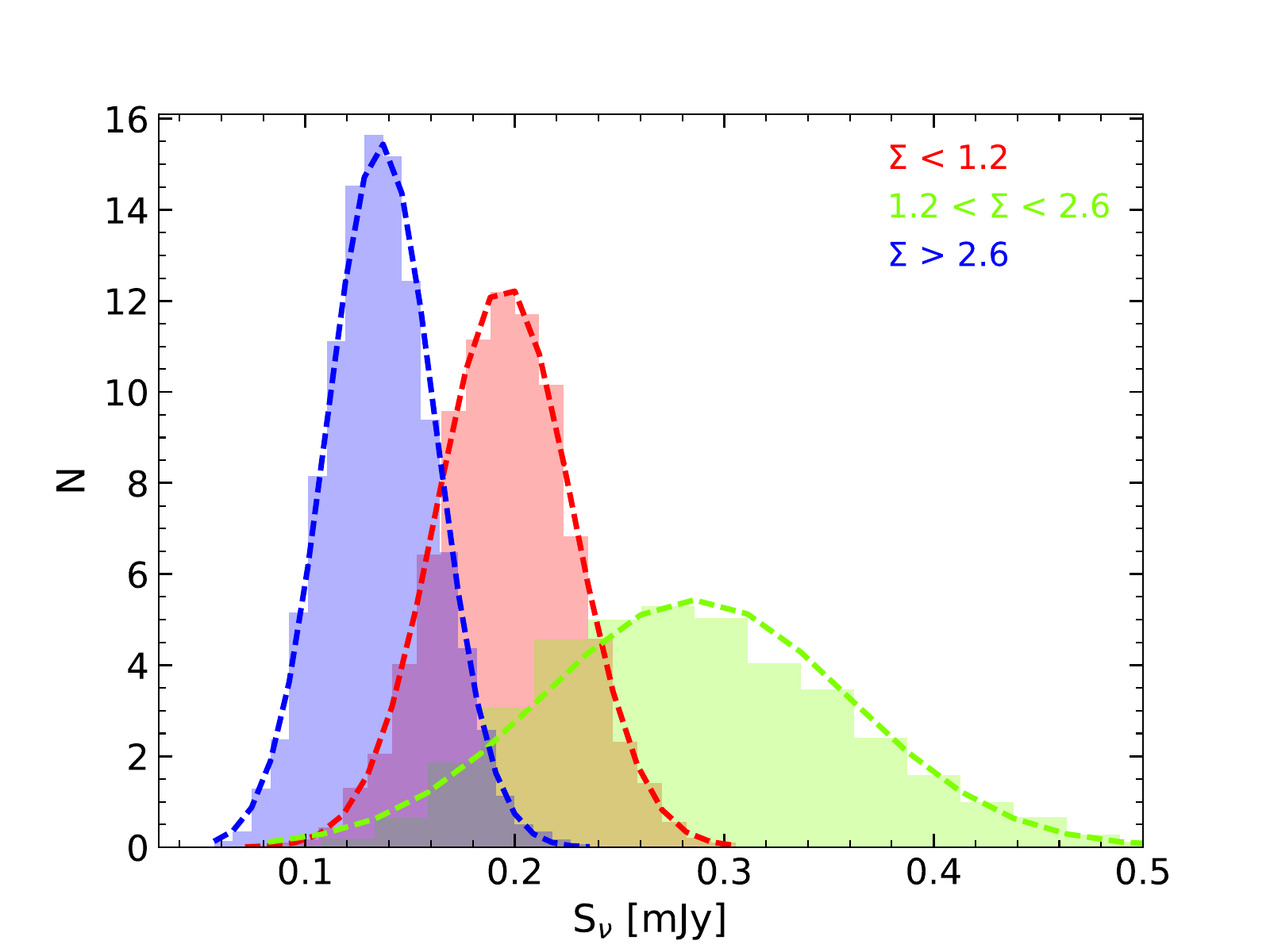}
\caption{Histograms of the weighted mean stacked integrate aperture flux measurements for $N=5000$ bootstrap realizations of sources in each of the three density bins. }
\label{bootstrap}
\end{figure} 

The local galaxy densities range from $0.16-10.5$ galaxies Mpc$^{-2}$.  In Figure \ref{boxcarStacks}, we plot a sliding box average of 8 samples; the sample separates into three 3 groups:  $0.16-1.2$, $1.2-2.6$, $2.6-10.5$ galaxies Mpc$^{-2}$.  These groups roughly correspond to field, filament, and cluster galaxies, respectively \citep[see][]{Darvish2017}.  In the top panel of Figure \ref{boxcarStacks}, there is a section of intermediate densities where the submm flux is elevated relative to low and high densities.  This persists when normalized by stellar mass (bottom panel).      

As the SGs were imaged such that their beamsizes are the same, both SGs can be easily stacked together.  The individual galaxies are stacked in two ways: a median stack and a weighed mean stack.  The mean stacked images were weighted by the square of the rms noise given by
\begin{equation}
S_{\mathrm{bin}} = \frac{\Sigma_{i=1}^{N_{bin}} S_i / \sigma^2_i}{\Sigma_{i=1}^{N_{bin}} 1 / \sigma^2_i},
\end{equation} 
where $S_{\mathrm{bin}}$ is the stacked flux density of N sources, $S_i$ is the flux density of each source and $\sigma_i$ is the rms noise of each source.  
Figure \ref{stacks} shows the weighted mean stack for the three local galaxy density bins.    

To confirm the size of the aperture, we again calculate the S/N for several aperture sizes in order to determine which aperture encloses the most signal relative to the noise.  We find the aperture radius of 0.54$\arcsec$ remains optimal. 
 
The median and weighted mean flux measurements are listed in Table \ref{StackedSamples} along with derived $M_\mathrm{ISM}$ and gas mass fractions.  The three stacked images are significantly detected in both the median and mean stacking.  $M_\mathrm{ISM}$ was calculated following equation \ref{MassISM}, where $z$, $\nu_\mathrm{obs}$, $d_L$, and $\Gamma_{\mathrm{RJ}}$ are the means of each bin.  

\subsection{Stacking Noise Estimates}
Noise estimates on the flux measurements for the stacking method are calculated two ways.  We first estimate the uncertainties on the weighted mean and median stacked continuum maps for each density bin using the method described in section \ref{FluxMeasurementSection}.  We also perform a bootstrap analysis to verify that a handful of sources are not biasing the stacks \citep[e.g.][]{Jauzac2011,Bethermin2012}.  We repeated the stacking process and total aperture flux measurement for $N = 5000$ realizations using randomly selected sources in each density bin with replacement.  The histograms of integrated aperture flux measurements are shown in Figure \ref{bootstrap}.  The bootstrap uncertainties correspond to the standard deviation of the 5000 realizations and are listed in Table \ref{StackedSamples}.  We find that the uncertainties from our bootstrap analysis are comparable with the uncertainties found initially using the stacked continuum maps, and therefore use the initial stacked uncertainties for all further analysis. 

\begin{deluxetable*}{lccccccccccc}[tb]
\centering
\tabletypesize{\scriptsize}
\tablecaption{Derived Parameters for Stacked Samples
\label{StackedSamples}
}
\tablehead{\colhead{Stack} & \colhead{RMS} &  \colhead{$S_\nu$} & \colhead{$\sigma_\nu$} & \colhead{$\sigma_{\nu, boot}$} & \colhead{S/N} & \colhead{$\langle z \rangle$} & \colhead{$\Sigma$} &  \colhead{$\langle$M$_*$$\rangle $} & \colhead{$\langle$SFR$_\mathrm{UV+IR}\rangle$} & \colhead{$\langle M_{\mathrm{ISM}} \rangle $} & \colhead{$\langle f_\mathrm{gas} \rangle$\tablenotemark{a}} \\
\colhead{} & \colhead{(mJy/beam)}  & \colhead{(mJy)} & \colhead{(mJy)} & \colhead{(mJy)} & \colhead{} & \colhead{} & \colhead{(Mpc$^{-2}$)} & \colhead{$(10^{10} M_{\odot})$} & \colhead{ ($M_\odot \mathrm{yr}^{-1}$)} & \colhead{$(10^{10} M_\odot)$} & \colhead{} 
}
\startdata	
\multicolumn{12}{c}{Low $\Sigma $ ($<$ 1.2 Mpc$^{-2})$ $-$ 61 galaxies}\\	
\hline																								 (Weighted) Mean	&	0.024& 0.201	&	0.019	& 0.032 &	10.9	&	0.742	&	0.6	&	3.83	$\pm$	0.39	&	38.35	$\pm$	4.04	&	0.78	$\pm$	0.07			&				0.17	$\pm$	0.02	\\			
Median	&	0.030 & 0.179	&	0.021	& 0.032 &	8.6	&	0.739	&	0.6	&	2.84	&	29.55	&	0.69		&				0.20	\\
\hline
\multicolumn{12}{c}{Intermediate $\Sigma $ (1.2 $-$ 2.6 Mpc$^{-2})$ $-$ 25 galaxies}\\	
\hline																								
(Weighted) Mean	& 0.035 & 	0.299	&	0.028	& 0.073 &	10.5	&	0.740	&	1.9	&	4.52	$\pm$	0.61	&	49.64	$\pm$	10.08	&	1.16	$\pm$	0.11	&				0.20	$\pm$	0.03	\\
Median	&	0.043 & 0.261	&	0.026	& 0.073 & 	10.1	&	0.733	&	1.9	&	3.47		&	33.45	&	1.01		&				0.23	\\
\hline
\multicolumn{12}{c}{High $\Sigma $ ($>$ 2 .6 Mpc$^{-2})$ $-$ 15 galaxies}\\
\hline																								
 (Weighted) Mean	& 0.047 & 	0.141	&	0.038& 0.025	&	3.76	&	0.740	&	6.3	&	3.93	$\pm$	0.97	&	28.26	$\pm$	1.99	&	0.55	$\pm$	0.15			&				0.12	$\pm$	0.04	\\
Median	&	0.058& 0.151	&	0.036	& 0.025&	4.26	&	0.733	&	6.3	&	2.69		&	27.51	&	0.58			&				0.17	\\
\enddata
\tablecomments{RMS, $S_\nu$, $\sigma_\nu$ are found from a weighted mean of the individual sources.  $\sigma_{\nu,boot}$ is found from a bootstrap method.  All others are either derived from $S_\nu$ or a normal mean. Uncertainties for $\langle$M$_*\rangle$ and $\langle$SFR$_\mathrm{UV+IR}\rangle$ are given as the standard error of the mean.} 
\tablenotetext{a}{$\langle f_\mathrm{gas}\rangle =\langle$M$_\mathrm{ISM}\rangle$/($\langle$M$_*\rangle$ + $\langle$M$_\mathrm{ISM}\rangle$)}
\end{deluxetable*}

\begin{figure}[tb]
    \begin{minipage}{.5\textwidth}
    \centering
    \includegraphics[width=0.95\linewidth]{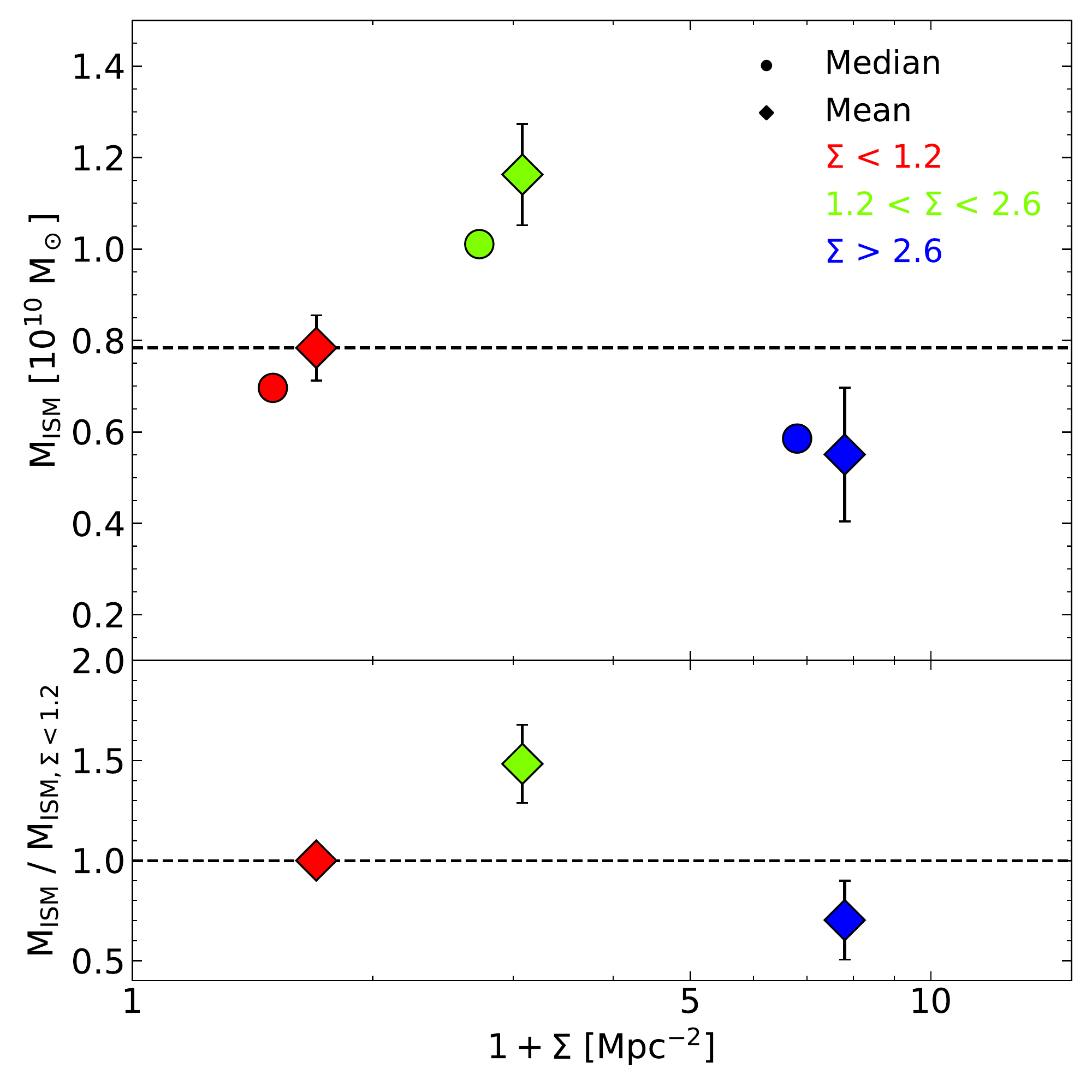}
    \caption{M$_{\rm ISM}$ (upper) and M$_{\rm ISM}$ normalized by the ISM Mass of the lowest density bin (lower) as a function of density.  Relative to low densities, the intermediate density bin has an increase in ISM mass which then drops off at higher densities.}
    \label{allDensityA}
    \end{minipage}%
    \begin{minipage}{0.5\textwidth}
    \centering
     \includegraphics[width=0.95\linewidth]{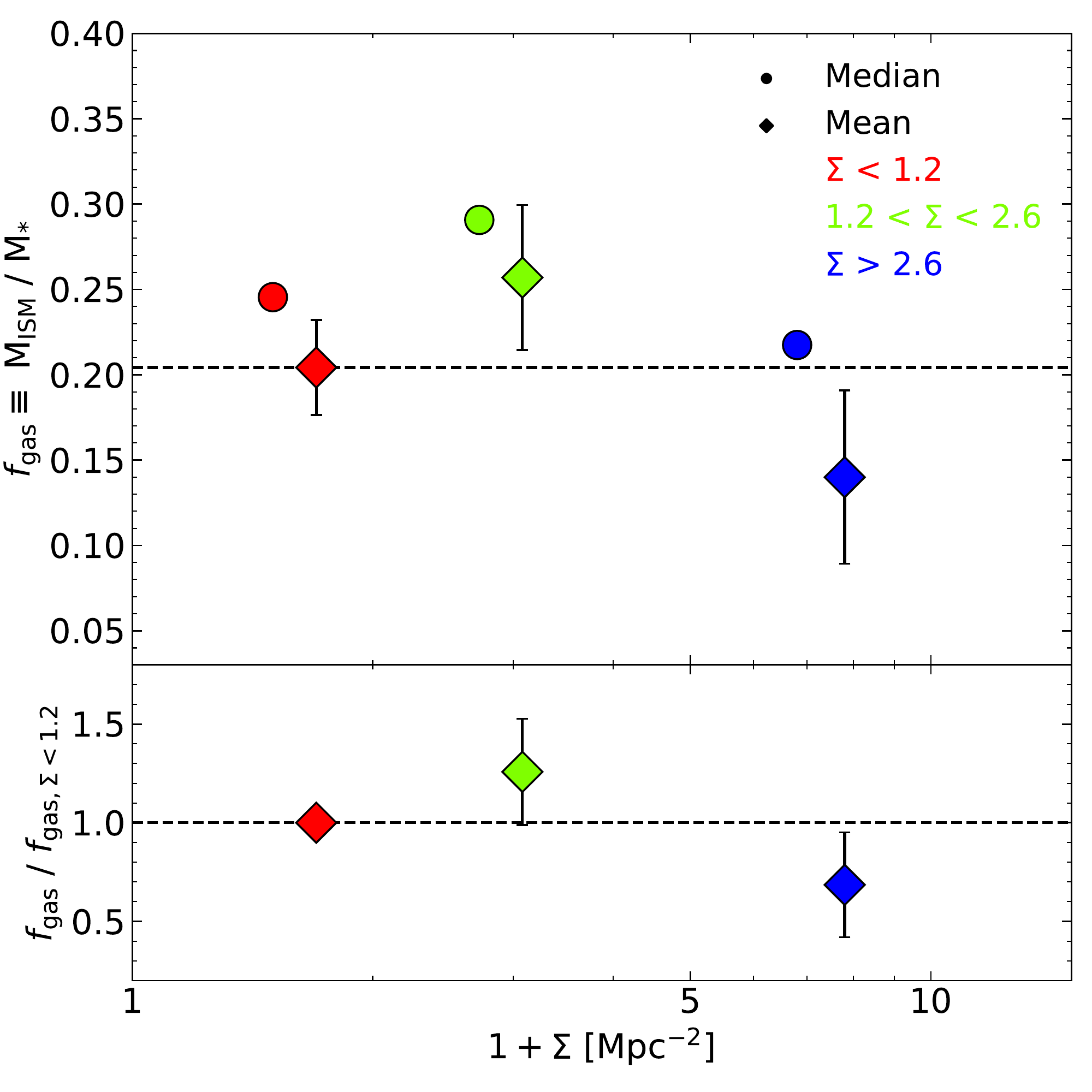}
     \caption{Gas mass fraction (f$_{\rm gas}$; upper) and f$_{\rm gas}$ normalized by the f$_{\rm gas}$ of the lowest density bin (lower) as a function of density. When normalized by the stellar mass, the trend in ISM mass shown in Figure \ref{allDensityA} becomes less significant. }
     \label{AllDensityB}
    \end{minipage}
\end{figure}

\begin{figure}
    \begin{minipage}{0.5\textwidth}
    \centering
    \includegraphics[width=0.95\linewidth]{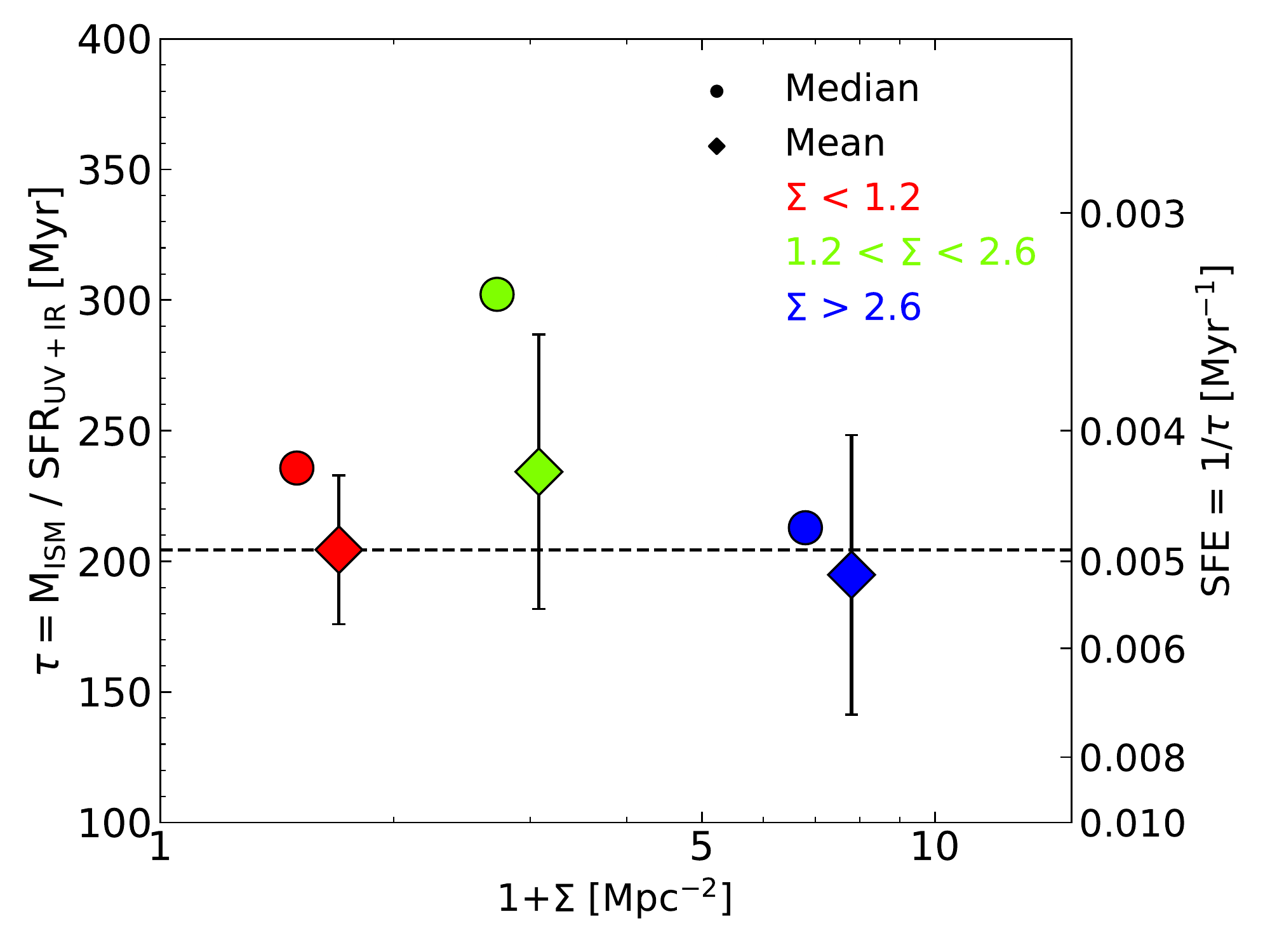}
    \caption{Depletion time (or SFE, right Y-axis) as a function of density. Given the uncertainties, the gas appears to be depleted at the same rate across all densities.}
    \label{fig:DepTime}
    \end{minipage}
\end{figure}

\section{Results}
\subsection{Individual detections}\label{5.1}
Flux and mass measurements for the detected sources along with 3$\sigma$ upper limits for non detections are given in Table \ref{INFO} along with SFE and gas mass fractions.  

Of the 68 detections, 31 are significant in total aperture flux while 37 are significant in peak pixel flux.  All total aperture flux detections are also detected in peak pixel flux.  If the total aperture flux is higher than the peak flux, this indicates that the source is marginally resolved.  Therefore, for these sources, we use the aperture flux.  From individual detections, we do not see any trends in flux density with galaxy environment, with both the integrated and peak pixel flux measurements showing large scatter over the range of densities.  We turn to stacking the galaxies in different density bins to look for trends with environment.  
    
\subsection{Stacking Results}

\subsubsection{ISM Mass and Gas Mass Fraction in different Environments} \label{ISM-Density}

Using the median and weighted mean stacks from Figure \ref{stacks}, we calculate M$_{\rm ISM}$ for all density bins.  We show local galaxy density as a function of M$_{\rm ISM}$ in the top panel of Figure \ref{allDensityA}.    
The intermediate (filament) density bin has an increase in ISM mass relative to the low (field) density bin, which then falls off at higher (cluster) densities.  As shown in the bottom panel of Figure \ref{allDensityA}, relative to the low density bin, the galaxies at intermediate densities have ISM masses higher by a factor of $1.5 \pm 0.2$, ($\gtrsim 2.5\sigma$ from 1) while galaxies in the highest density bin have ISM masses lower by a factor of $0.7 \pm 0.2$ ($\sim 1.5\sigma$ from 1).  Between the intermediate and high density, the ISM mass decreases by a factor of $2.1 \pm 0.6$ ($\sim 2\sigma$).

To further assess the significance of the result that the ISM masses change with environment, we perform a 2D Anderson-Darling statistical (AD) tests to calculate the probability that the mm fluxes (and thus ISM masses) of galaxies in our three density bins are drawn from the same distribution. We test the different fluxes between the low/high, the low/intermediate, and the intermediate/high density bins.  The critical values for the significance levels [25$\%$, 10$\%$, 5$\%$, 2.5$\%$, 1$\%$] are [0.325, 1.226, 1.961, 2.718, 3.752].  The mm fluxes of galaxies in the low/high and low/mid density bins have a non-negligible probability of coming from the same distribution.  However, we find a AD statistic of 2.5 between the mm fluxes in the intermediate and high densities bins which indicates there is a low probability ($\sim9\%$) these densities come from the same distribution suggesting the different ISM masses between galaxies in intermediate and high densities may be a robust environmental effect.     

As we show in Figures \ref{histograms} and \ref{SFRandsSFR}, the stellar mass was selected to be roughly consistent across the sample, though the intermediate density stack has a slightly higher average stellar mass.  The gas fraction ($f_\mathrm{gas} = M_\mathrm{ISM} / (M_\mathrm{ISM} + M_{*})$) shown in Figure \ref{AllDensityB} follows a similar trend as $M_\mathrm{ISM}$ with density, though the uncertainties in stellar mass decrease the significance of this trend.  The large stellar mass uncertainties are based on SED fitting and are dependent on the SED model templates \citep[see][]{Laigle2016}.  Regardless of density, the $f_{\rm gas}$ values we calculate are consistent with estimates from \citet{Scoville2014}, \citet{Scoville2017}, and \cite{Tacconi2013} accounting for differences in mass ranges, and whether the estimates were found from only CO/dust detections or also include nondetections.  Given the uncertainties in $f_{\rm gas}$ along with no significant dependence of $f_{\rm gas}$ with density, we will focus on the effect of ISM mass on environment and its role in driving galaxy evolution.


\subsubsection{Depletion time in different Environments}\label{GMFenviro}
As shown in Figure \ref{fig:DepTime}, the depletion time ($\tau=1/ {\rm SFE} =  \mathrm{M}_\mathrm{ISM} / \mathrm{SFR}$) is relatively constant with density, $\tau = 210 \pm 71$ Myr, since the ISM mass increases with the star formation rate in our sample. 
Though the ISM gas is higher in galaxies in intermediate densities compared to lower or higher densities, galaxies in these environments use their ISM faster and it is depleted on the same timescale across all environments. 

The depletion timescales measured in our sample ($\sim~210$ Myr) are lower than estimated by \citet{Tacconi2018}, where the relation between depletion timescale and redshift is proposed to be $\tau \sim (1+z)^{-0.62\pm0.13}$ which gives $\tau \sim 0.72$ Gyr at $z\sim0.7$. However, the \citet{Tacconi2018} relation is for galaxies on the MS and our sample lies slightly above the MS where lower depletion timescales are expected due to an increase in SFR \citep{Kennicutt2012, Saintonge2012}.


\section{Discussion} \label{Discussion}

Given the ISM masses of galaxies in the intermediate and high density bins are inconsistent with those at the low densities (bottom panel of Figure \ref{allDensityA}) and there is a drop in ISM mass from intermediate density to low densities, our data suggests that there is a dependence of the ISM mass on environment. The increase in ISM mass in galaxies at intermediate densities might suggest an increase in mergers or interactions that drive more molecular gas into galaxies. 
The drop in the ISM mass between the intermediate and high densities might indicate that we are catching galaxies at this epoch where the environmental effects are beginning to take effect and galaxies in high density environments are beginning to lose their gas.  As the high density bin also has a slightly lower SFR compared to the low and intermediate bins, the drop in both star formation and ISM mass could potentially be due to environmental processes which remove gas, such as ram pressure stripping \citep[RPS;][]{Gunn1972} or strangulation \citep{Larson1980}, and quench the star formation in high density environments.  These results are suggestive and more observations of the ISM in intermediate and high density galaxies are needed to test these ideas.

\subsection{Comparison to other studies}
With a sample of 708 high redshift galaxies with ALMA dust continuum measurements, \citet{Scoville2017} found strong dependencies between the ISM mass of a galaxy with redshift and distance from the MS, but they did not investigate the effects of environment.  
A recent study by \citet{Darvish2018b} did not find any dependence of the gas mass fraction and depletion timescales on environment using the \citet{Scoville2017} sample.  However, given that their sample did not probe a large dynamic range of environments at any given redshift, it is difficult to separate any evolution with environment from the known strong evolution with redshift and sSFR.  In this study, we focused on a single redshift with a known LSS and a large range of environments, and found that the environment does play a role in how the gas is used up in different densities relative to the known evolution with redshift and distance from the MS.   

We find that in low (field) and intermediate (filament) densities, the environment does not seem to affect the depletion time or the gas mass fraction of the galaxies, in agreement with \citet{Darvish2018b}.  However, at high density, the decrease in the ISM mass at $2\sigma$ significance from the intermediate density, and therefore gas mass fraction, indicates the environment does have an influence on the evolution of a galaxy.  As a reminder, this sample was selected to include all sources at this redshift interval above $\log(\mathrm{M}_*/\mathrm{M}_\odot) > 9.5$, with far-IR detections to ensure a submillimeter detection with ALMA.  This sample should be representative as a function of density, since we do not find a strong dependency on the fraction of star formation that is obscured by dust with density.

\subsection{Comparing the evolution of ISM mass with density and redshift}
We now investigate how the rate of evolution of ISM mass with density compares to the known evolution in ISM mass with redshift.  From Figure \ref{AllDensityB}, there is a factor of 1.8 between the observed gas fraction at intermediate densities (f$_{\rm gas}$ = 0.256) and high densities (f$_{\rm gas}$  = 0.139). 

From CO measurements, it is well known that f$_{\rm gas}$ increases with increasing redshift; f$_{\rm gas}$ = M$_\mathrm{ISM}/$M$_* \sim 0.1 \times (1+z)^2$ (see Figure 9 from \citet{Carilli2013}).  Recently, \citet{Tacconi2018} combined recent gas mass fractions from CO flux lines, far IR dust spectral energy distributions, and dust continuum to determine a new scaling relation for gas fraction of galaxies between $z\sim 0-4$.  Using the observed relations between gas fraction and redshift from \citet{Tacconi2018} and \citet{Carilli2013}, we find that at $z=0.73$ it takes a galaxy $2-3$ Gyrs to experience a decrease in the gas fraction by a factor of 1.8, comparable to what we find between the intermediate and high density bins at a single redshift. This is a long timescale compared to the depletion time of the gas ($\sim$200 Myr) and indicates that, at least at high densities, the environment may be driving the decrease in the molecular gas and thus the star formation. 

\subsection{Role of Environment in Gas Depletion}
\subsubsection{Ram Pressure Stripping}
We consider our intermediate density bin to be representative of filament galaxies and our highest density bin to be representative of cluster galaxies (\citep[e.g.][]{Darvish2017}).  Most of these high density galaxies fall within the cluster at $z\sim0.7$ detected by \citet{Scoville2007b} and \citet{Guzzo2007}.  This cluster was originally found by adaptive smoothing of galaxy counts from photometric redshift catalogues \citep{Scoville2007b}.  \citet{Guzzo2007} used follow up weak lensing and X-ray observations to calculate a cluster mass $> 10^{14}\ {\rm M}_\odot$ suggesting that the cluster is a true virialized structure.  Therefore, as stated in section \ref{ISM-Density}, an explanation for the depletion of ISM mass at high densities could be due to RPS, in which the hot inter cluster medium (ICM) removes the ISM mass of a galaxy, or strangulation, when the ICM removes the hot halo of a cluster and does not allow refueling over several Gyrs.    

Hydrodynamical simulations of RPS of individual galaxies with the RPS estimation from \citet{Gunn1972} has found that gas can be removed within $\sim 10-200$ Myr \citep{Abadi1999, Marcolini2003, Roediger2006, Roediger2007, Kronberger2008, Steinhauser2016}.  On the other hand, strangulation, which prevents accretion of fresh cold gas in the hot ICM, occurs on Gyr time-scales and results in the SFR of cluster galaxies to decrease to levels consistent with field galaxies, which is seen in our highest density bin.  Both of these physical processes  can help explain the decrease in ISM mass and SFR seen in our high density sample.  
This indicates that in cluster environments, cold gas from galaxies is begin stripped away quickly resulting in lower SFRs and mass while also allowing for the dust to be heated and obscured. 

\subsubsection{Mergers and Morphology}
In these filamentary-cluster environments, an increased rate of mergers relative to the field environments could explain the rise in M$_{\rm ISM}$ relative to the field galaxies. With an increased merger rate in higher density environments, the gas in these galaxies can be stripped, heated, or efficiently funnel gas to increase the SFR. Using the visual morphological classifications from \citet{Kartaltepe2015} for our galaxies, we find that there is a small increase in the number of mergers at intermediate densities ($44\pm15\%$ of sources) compared to low ($36\pm8\%$ of sources) and high ($20\pm12\%$ of sources) density environments, though not statistically significant.  Therefore, though it is possible that galaxies in filaments have a higher rate of interacting leading to the increase in measured ISM mass, larger samples are required to test this.

Further observations of galaxies at predominantly intermediate and high densities in this field at $z\sim0.7$ will help confirm the increased gas depletion and decreased ISM mass at the highest densities.  Observations of surrounding redshifts along the $z\sim 0.7$ LSS will help provide further evidence that environmental factors such as mergers and RPS affect the evolution on timescales that cannot be accounted for by redshift evolution.


\section{Conclusion}
In this paper, we explore the role of environment on star formation by looking at the ISM gas mass, depletion timescales, and gas mass fraction as a function of density. We look at a sample of 101 galaxies at $z\sim0.7$ in the COSMOS 2 deg$^2$ survey over a range of local galaxy densities ($0.16 < \Sigma < 10.5$ Mpc$^{-2}$) with Band 7 observations with ALMA.  We use dust continuum to probe the ISM content and stack the galaxies by density bins in order to probe the overall trends of environment at this epoch when star formation starts to depend on density.  
We find:
\begin{enumerate}
\item ISM masses are individually detected in 68 galaxies between $z = 0.72-0.76$ in a range of environments.   These galaxies are all on or slightly above the MS and have consistent average SFRs and M$_*$ as a function of environment.              
\item We stack the galaxies into three density bins (low/intermediate/high corresponding roughly to field/filament/cluster). Relative to galaxies in the the low (field) density bin, we find elevated submm flux and ISM mass in galaxies in the intermediate (filament) density bin (at $2.5\sigma$ significance) and lower values for galaxies in the high (cluster) density bin (at $1.5\sigma$ significance).  At $2\sigma$ significance, there is a decrease in ISM mass content in galaxies from intermediate to high density environments. 
\item The gas depletion timescales are relatively constant across all environments.  
\item At this specific redshift, the environment at high densities is affecting the gas supply, as the drop in gas fraction from intermediate to high densities would take $\sim2-3$ Gyr to occur without environmental influences. 
\end{enumerate}
These results suggest that at this critical epoch, intermediate filament environments can potentially provide the optimal conditions to continue star formation, while environments that are more dense start to quench their star formation.
Mergers and environmental processes such as RPS and strangulation together regulate the gas available to form stars in galaxies in different environments.  Additional observations that focus at a single redshift but a broad range of environments will help to confirm these results.   

\acknowledgments
We thank the anonymous referee for the suggestions and comments that have significantly improved this manuscript.  S.K.B. and A.P. acknowledge support from NASA ADAP 13-0054.  S.K.B. is thankful for the North American ALMA Science Center support staff, especially Erica Keller, for all their help with data reduction.  This paper makes use of the following ALMA data: ADS/JAO.ALMA $\#$2015.1.00055.S. ALMA is a partnership of ESO (representing its member states), NSF (USA) and NINS (Japan), together with NRC (Canada), MOST and ASIAA (Taiwan), and KASI (Republic of Korea), in cooperation with the Republic of Chile. The Joint ALMA Observatory is operated by ESO, AUI/NRAO and NAOJ.  The National Radio Astronomy Observatory is a facility of the National Science Foundation operated under cooperative agreement by Associated Universities, Inc.  
   
\facilities{ALMA, HST, {\it Herschel}, {\it Spitzer}}
\software{astropy \citep{Astropy-CollaborationRobitaille:2013}, CASA \citep{McMullin2007}}

\singlespace
													
\begin{longrotatetable}
\begin{deluxetable*}{cccccccccccccccc}
\tabletypesize{\scriptsize}
\setlength{\tabcolsep}{0.06in}
\tablecaption{Sample properties and Measurements
\label{INFO}
}
\tablehead{\colhead{ID} & \colhead{SG \tablenotemark{a}}&\colhead{R.A. (J2000)} &  \colhead{Dec (J2000)} &\colhead{z$_\mathrm{spec}$} & \colhead{$\Sigma$} & \colhead{SFR$_\mathrm{IR}$} & \colhead{SFR$_\mathrm{UV}$} & \colhead{SFR$_\mathrm{total}$} &\colhead{Log(M$_{*}$)}  & \colhead{RMS} &  \colhead{S$_\nu$} &\colhead{$M_{\mathrm{ISM}}$} & \colhead{Log(SFE)} & \colhead{$f_\mathrm{gas}$\tablenotemark{b}} & \colhead{Notes\tablenotemark{c}} \\ \colhead{} & \colhead{} & \colhead{(deg)}& \colhead{(deg)} & \colhead{} & \colhead{(Mpc$^{-2}$)} &  \colhead{($M_\odot$/yr)} &  \colhead{($M_\odot$/yr)} &  \colhead{($M_\odot$/yr)} &\colhead{(M$_\odot$)}  & \colhead{(mJy/beam)} & \colhead{(mJy)}  & \colhead{(10$^{10}$ M$_\odot$)} &  \colhead{(yr$^{-1}$)} &  \colhead{} & \colhead{}
}
\startdata
\multicolumn{16}{c}{Bin 1: Low $\Sigma $ ($<$1.2 Mpc$^{-2})$ $-$ 61 galaxies}\\
\hline													
\hline									
65511	&	2	&	150.76950070	&	2.44000006	&	0.746	&	0.17	&	44.61	&	1.34	&	45.95	&	10.69	&	0.216	&	0.38	$\pm$	0.15	&	1.52	$\pm$	0.59	&	-8.52	$\pm$	0.17	&	0.24	$\pm$	0.10	&	1	\\
12957	&	2	&	149.90295410	&	1.98539996	&	0.744	&	0.19	&	21.12	&	1.57	&	22.69	&	10.75	&	0.214	&	0.44	$\pm$	0.15	&	1.75	$\pm$	0.58	&	-8.89	$\pm$	0.14	&	0.24	$\pm$	0.08	&	2	\\
2121	&	1	&	149.59567260	&	1.92436004	&	0.728	&	0.20	&	60.07	&	0.52	&	60.59	&	10.66	&	0.177	&	0.62	$\pm$	0.11	&	2.35	$\pm$	0.42	&	-8.59	$\pm$	0.08	&	0.34	$\pm$	0.07	&	2	\\
3095	&	1	&	149.64450070	&	1.82980001	&	0.738	&	0.23	&	17.89	&	0.48	&	18.37	&	10.80	&	0.174	&	0.44	$\pm$	0.10	&	1.67	$\pm$	0.39	&	-8.96	$\pm$	0.10	&	0.21	$\pm$	0.05	&	2	\\
33697	&	2	&	150.12416080	&	2.43136001	&	0.757	&	0.26	&	18.80	&	2.33	&	21.13	&	10.32	&	0.213	&	$<$	0.53	&	$<$	2.14	&	$<$	-9.01	&	$<$	0.51	&	3	\\				
19710	&	2	&	150.02076720	&	2.42260003	&	0.758	&	0.29	&	46.36	&	3.95	&	50.31	&	10.45	&	0.218	&	$<$	0.39	&	$<$	1.58	&	$<$	-8.50	&	$<$	0.36	&	3	\\				
59452	&	2	&	150.50376890	&	1.85462999	&	0.746	&	0.32	&	30.30	&	1.22	&	31.53	&	10.40	&	0.210	&	0.45	$\pm$	0.14	&	1.80	$\pm$	0.57	&	-8.76	$\pm$	0.14	&	0.42	$\pm$	0.14	&	2	\\
50059	&	1	&	150.28323360	&	2.10188007	&	0.724	&	0.33	&	18.87	&	2.18	&	21.05	&	9.85	&	0.167	&	$<$	0.37	&	$<$	1.40	&	$<$	-8.82	&	$<$	0.66	&	3	\\				
48162	&	1	&	150.24334720	&	2.27336001	&	0.740	&	0.33	&	20.27	&	0.58	&	20.85	&	10.85	&	0.163	&	0.28	$\pm$	0.12	&	1.09	$\pm$	0.45	&	-8.72	$\pm$	0.18	&	0.13	$\pm$	0.06	&	1	\\
1576	&	2	&	149.56404110	&	2.23825002	&	0.748	&	0.34	&	28.21	&	1.87	&	30.07	&	10.49	&	0.214	&	0.33	$\pm$	0.15	&	1.30	$\pm$	0.58	&	-8.64	$\pm$	0.19	&	0.30	$\pm$	0.14	&	1	\\
22698	&	1	&	150.06425480	&	1.93305004	&	0.738	&	0.36	&	20.71	&	4.97	&	25.68	&	9.95	&	0.161	&	0.53	$\pm$	0.10	&	2.04	$\pm$	0.40	&	-8.90	$\pm$	0.09	&	0.70	$\pm$	0.17	&	2	\\
4377	&	1	&	149.69613650	&	2.46111989	&	0.732	&	0.39	&	19.33	&	2.80	&	22.13	&	9.80	&	0.176	&	$<$	0.43	&	$<$	1.62	&	$<$	-8.86	&	$<$	0.72	&	3	\\				
13692	&	1	&	149.91577150	&	1.67797005	&	0.740	&	0.40	&	32.25	&	0.22	&	32.47	&	10.70	&	0.173	&	0.60	$\pm$	0.12	&	2.29	$\pm$	0.44	&	-8.85	$\pm$	0.08	&	0.31	$\pm$	0.07	&	1	\\
30358	&	2	&	150.10627750	&	1.70779002	&	0.745	&	0.41	&	22.89	&	1.36	&	24.24	&	10.59	&	0.211	&	0.37	$\pm$	0.14	&	1.49	$\pm$	0.57	&	-8.79	$\pm$	0.17	&	0.28	$\pm$	0.11	&	1	\\
2331	&	1	&	149.60787960	&	2.60363007	&	0.738	&	0.41	&	57.72	&	2.62	&	60.34	&	10.56	&	0.180	&	$<$	0.40	&	$<$	1.52	&	$<$	-8.40	&	$<$	0.30	&	3	\\				
55106	&	2	&	150.38865660	&	2.31001997	&	0.744	&	0.42	&	28.94	&	0.93	&	29.87	&	9.83	&	0.215	&	0.56	$\pm$	0.15	&	2.23	$\pm$	0.60	&	-8.87	$\pm$	0.12	&	0.77	$\pm$	0.26	&	2	\\
38620	&	1	&	150.14944460	&	1.92332006	&	0.731	&	0.43	&	17.60	&	1.87	&	19.46	&	10.40	&	0.166	&	$<$	0.36	&	$<$	1.37	&	$<$	-8.85	&	$<$	0.35	&	3	\\				
15068	&	2	&	149.94177250	&	2.39576006	&	0.758	&	0.43	&	74.73	&	2.37	&	77.10	&	10.94	&	0.230	&	0.61	$\pm$	0.18	&	2.45	$\pm$	0.72	&	-8.50	$\pm$	0.13	&	0.22	$\pm$	0.07	&	1	\\
31543	&	2	&	150.11227420	&	2.38735008	&	0.757	&	0.44	&	33.06	&	1.12	&	34.19	&	10.25	&	0.216	&	0.48	$\pm$	0.13	&	1.92	$\pm$	0.54	&	-8.75	$\pm$	0.12	&	0.52	$\pm$	0.16	&	2	\\
3358	&	2	&	149.65591430	&	1.92480004	&	0.748	&	0.45	&	25.40	&	6.15	&	31.55	&	9.96	&	0.211	&	$<$	0.42	&	$<$	1.68	&	$<$	-8.73	&	$<$	0.65	&	3	\\				
3972	&	1	&	149.68240360	&	2.19365001	&	0.734	&	0.46	&	22.90	&	1.38	&	24.27	&	10.41	&	0.186	&	$<$	0.41	&	$<$	1.57	&	$<$	-8.81	&	$<$	0.38	&	3	\\				
32787	&	2	&	150.11898800	&	2.32452011	&	0.756	&	0.48	&	25.56	&	0.00	&	25.56	&	11.05	&	0.203	&	0.50	$\pm$	0.14	&	2.00	$\pm$	0.55	&	-8.89	$\pm$	0.12	&	0.15	$\pm$	0.04	&	2	\\
16555	&	1	&	149.96690370	&	1.69801998	&	0.737	&	0.50	&	26.38	&	0.87	&	27.25	&	10.56	&	0.164	&	0.33	$\pm$	0.10	&	1.25	$\pm$	0.37	&	-8.66	$\pm$	0.13	&	0.26	$\pm$	0.08	&	2	\\
57520	&	1	&	150.45071410	&	2.01388001	&	0.729	&	0.51	&	18.03	&	4.16	&	22.19	&	10.49	&	0.161	&	$<$	0.35	&	$<$	1.32	&	$<$	-8.77	&	$<$	0.30	&	3	\\				
44291	&	1	&	150.18431090	&	2.42952991	&	0.739	&	0.54	&	19.96	&	1.53	&	21.49	&	9.97	&	0.162	&	$<$	0.35	&	$<$	1.33	&	$<$	-8.79	&	$<$	0.59	&	3	\\				
54987	&	1	&	150.38655090	&	1.72935998	&	0.736	&	0.56	&	17.87	&	2.20	&	20.07	&	10.65	&	0.165	&	0.37	$\pm$	0.10	&	1.41	$\pm$	0.39	&	-8.85	$\pm$	0.12	&	0.24	$\pm$	0.07	&	2	\\
40883	&	1	&	150.16107180	&	2.27872992	&	0.725	&	0.57	&	45.39	&	0.38	&	45.76	&	10.55	&	0.164	&	0.39	$\pm$	0.10	&	1.48	$\pm$	0.39	&	-8.51	$\pm$	0.11	&	0.29	$\pm$	0.08	&	2	\\
51707	&	2	&	150.32321170	&	2.44926000	&	0.744	&	0.57	&	26.04	&	2.48	&	28.52	&	10.30	&	0.221	&	$<$	0.43	&	$<$	1.70	&	$<$	-8.78	&	$<$	0.46	&	3	\\				
57753	&	2	&	150.45622250	&	2.63929009	&	0.745	&	0.58	&	19.26	&	0.79	&	20.05	&	10.81	&	0.216	&	0.56	$\pm$	0.14	&	2.25	$\pm$	0.56	&	-9.05	$\pm$	0.11	&	0.26	$\pm$	0.07	&	1	\\
1460	&	1	&	149.55584720	&	1.79781997	&	0.739	&	0.59	&	19.61	&	0.69	&	20.29	&	10.33	&	0.163	&	0.33	$\pm$	0.10	&	1.25	$\pm$	0.40	&	-8.79	$\pm$	0.14	&	0.37	$\pm$	0.13	&	2	\\
5708	&	1	&	149.73233030	&	2.48087001	&	0.735	&	0.59	&	26.30	&	4.25	&	30.54	&	10.00	&	0.180	&	0.38	$\pm$	0.11	&	1.44	$\pm$	0.43	&	-8.67	$\pm$	0.13	&	0.59	$\pm$	0.21	&	2	\\
38196	&	1	&	150.14721680	&	2.33724999	&	0.727	&	0.60	&	49.17	&	2.65	&	51.83	&	10.86	&	0.163	&	0.65	$\pm$	0.12	&	2.46	$\pm$	0.46	&	-8.68	$\pm$	0.08	&	0.25	$\pm$	0.05	&	1	\\
57025	&	1	&	150.43638610	&	1.81391001	&	0.737	&	0.61	&	22.45	&	2.69	&	25.14	&	10.09	&	0.169	&	$<$	0.36	&	$<$	1.37	&	$<$	-8.74	&	$<$	0.53	&	3	\\				
42610	&	2	&	150.17243960	&	2.63491988	&	0.752	&	0.62	&	28.19	&	6.39	&	34.58	&	10.27	&	0.218	&	$<$	0.45	&	$<$	1.82	&	$<$	-8.72	&	$<$	0.49	&	3	\\				
19493	&	2	&	150.01629640	&	2.38214993	&	0.757	&	0.65	&	20.32	&	1.62	&	21.94	&	10.55	&	0.223	&	$<$	0.42	&	$<$	1.69	&	$<$	-8.89	&	$<$	0.32	&	3	\\				
10336	&	2	&	149.85395810	&	1.78397000	&	0.747	&	0.66	&	36.13	&	4.59	&	40.73	&	10.45	&	0.205	&	$<$	0.43	&	$<$	1.70	&	$<$	-8.62	&	$<$	0.38	&	3	\\				
9407	&	2	&	149.83442690	&	1.79604006	&	0.744	&	0.68	&	51.08	&	1.74	&	52.82	&	10.93	&	0.217	&	$<$	0.41	&	$<$	1.62	&	$<$	-8.49	&	$<$	0.16	&	3	\\				
17117	&	1	&	149.97621160	&	2.60859990	&	0.733	&	0.70	&	30.72	&	0.34	&	31.06	&	11.10	&	0.173	&	0.33	$\pm$	0.13	&	1.27	$\pm$	0.48	&	-8.61	$\pm$	0.17	&	0.09	$\pm$	0.04	&	1	\\
42231	&	1	&	150.16955570	&	1.83463001	&	0.740	&	0.70	&	34.97	&	1.28	&	36.25	&	10.46	&	0.170	&	0.26	$\pm$	0.12	&	0.99	$\pm$	0.45	&	-8.44	$\pm$	0.20	&	0.26	$\pm$	0.12	&	1	\\
23947	&	1	&	150.07217410	&	2.50096989	&	0.733	&	0.70	&	40.83	&	1.22	&	42.05	&	10.37	&	0.169	&	0.40	$\pm$	0.13	&	1.51	$\pm$	0.49	&	-8.55	$\pm$	0.14	&	0.39	$\pm$	0.14	&	1	\\
16561	&	2	&	149.96705630	&	1.84922004	&	0.752	&	0.71	&	23.05	&	2.68	&	25.73	&	11.10	&	0.204	&	$<$	0.46	&	$<$	1.86	&	$<$	-8.86	&	$<$	0.13	&	3	\\				
41590	&	1	&	150.16532900	&	2.28944993	&	0.728	&	0.75	&	26.98	&	0.53	&	27.50	&	10.28	&	0.165	&	0.44	$\pm$	0.10	&	1.67	$\pm$	0.37	&	-8.78	$\pm$	0.10	&	0.47	$\pm$	0.12	&	2	\\
34907	&	1	&	150.13040160	&	2.51457000	&	0.732	&	0.75	&	38.49	&	0.39	&	38.88	&	10.29	&	0.169	&	$<$	0.33	&	$<$	1.25	&	$<$	-8.51	&	$<$	0.39	&	3	\\				
14139	&	1	&	149.92431640	&	2.46900010	&	0.739	&	0.76	&	32.42	&	0.85	&	33.27	&	10.31	&	0.170	&	0.26	$\pm$	0.13	&	1.00	$\pm$	0.49	&	-8.48	$\pm$	0.21	&	0.33	$\pm$	0.17	&	1	\\
48244	&	1	&	150.24497990	&	2.22392011	&	0.727	&	0.77	&	18.63	&	3.49	&	22.12	&	9.85	&	0.167	&	$<$	0.35	&	$<$	1.31	&	$<$	-8.77	&	$<$	0.65	&	3	\\				
48525	&	1	&	150.25051880	&	1.98089004	&	0.723	&	0.86	&	42.54	&	1.34	&	43.88	&	10.25	&	0.173	&	$<$	0.36	&	$<$	1.34	&	$<$	-8.49	&	$<$	0.43	&	3	\\				
27057	&	1	&	150.09062200	&	2.41316009	&	0.727	&	0.87	&	17.95	&	1.19	&	19.14	&	10.21	&	0.164	&	0.33	$\pm$	0.10	&	1.26	$\pm$	0.36	&	-8.82	$\pm$	0.12	&	0.44	$\pm$	0.14	&	2	\\
58803	&	2	&	150.48532100	&	2.66752005	&	0.744	&	0.90	&	58.06	&	5.46	&	63.51	&	10.95	&	0.240	&	0.51	$\pm$	0.17	&	2.03	$\pm$	0.67	&	-8.51	$\pm$	0.14	&	0.19	$\pm$	0.06	&	2	\\
4433	&	2	&	149.69822690	&	2.36782002	&	0.752	&	0.90	&	23.17	&	0.83	&	24.00	&	9.83	&	0.206	&	0.46	$\pm$	0.14	&	1.86	$\pm$	0.57	&	-8.89	$\pm$	0.13	&	0.73	$\pm$	0.28	&	2	\\
21261	&	1	&	150.05067440	&	2.47757006	&	0.737	&	0.92	&	34.00	&	0.80	&	34.79	&	10.08	&	0.165	&	$<$	0.40	&	$<$	1.52	&	$<$	-8.64	&	$<$	0.56	&	3	\\				
46272	&	1	&	150.20808410	&	2.04952002	&	0.731	&	0.94	&	34.04	&	0.14	&	34.18	&	10.71	&	0.163	&	0.35	$\pm$	0.14	&	1.31	$\pm$	0.53	&	-8.58	$\pm$	0.18	&	0.20	$\pm$	0.09	&	1	\\
7093	&	1	&	149.77285770	&	2.55573010	&	0.735	&	0.96	&	30.25	&	6.53	&	36.79	&	10.10	&	0.178	&	0.44	$\pm$	0.12	&	1.69	$\pm$	0.45	&	-8.66	$\pm$	0.12	&	0.57	$\pm$	0.18	&	2	\\
51675	&	2	&	150.32260130	&	2.51247001	&	0.755	&	0.97	&	26.81	&	2.28	&	29.09	&	10.45	&	0.213	&	$<$	0.42	&	$<$	1.70	&	$<$	-8.77	&	$<$	0.38	&	3	\\				
21403	&	1	&	150.05282590	&	2.24354005	&	0.725	&	0.98	&	54.91	&	0.36	&	55.27	&	11.08	&	0.172	&	0.48	$\pm$	0.12	&	1.82	$\pm$	0.46	&	-8.52	$\pm$	0.11	&	0.13	$\pm$	0.04	&	1	\\
31361	&	2	&	150.11178590	&	2.58644009	&	0.755	&	0.99	&	46.23	&	1.36	&	47.59	&	10.60	&	0.215	&	0.60	$\pm$	0.13	&	2.41	$\pm$	0.54	&	-8.70	$\pm$	0.10	&	0.38	$\pm$	0.09	&	1	\\
46016	&	2	&	150.20339970	&	1.90285003	&	0.753	&	1.05	&	83.95	&	1.27	&	85.22	&	10.60	&	0.216	&	$<$	0.41	&	$<$	1.64	&	$<$	-8.28	&	$<$	0.29	&	3	\\				
48710	&	1	&	150.25398250	&	1.94953001	&	0.737	&	1.10	&	31.18	&	1.27	&	32.45	&	9.98	&	0.179	&	$<$	0.35	&	$<$	1.35	&	$<$	-8.62	&	$<$	0.58	&	3	\\				
24389	&	2	&	150.07476810	&	1.65954995	&	0.748	&	1.13	&	18.26	&	0.66	&	18.91	&	10.68	&	0.214	&	0.32	$\pm$	0.14	&	1.27	$\pm$	0.55	&	-8.83	$\pm$	0.19	&	0.21	$\pm$	0.09	&	1	\\
39871	&	1	&	150.15560910	&	2.78765011	&	0.726	&	1.14	&	18.45	&	0.50	&	18.95	&	10.45	&	0.182	&	$<$	0.37	&	$<$	1.41	&	$<$	-8.87	&	$<$	0.33	&	3	\\				
7147	&	1	&	149.77459720	&	2.47119999	&	0.735	&	1.18	&	98.91	&	0.26	&	99.16	&	10.75	&	0.181	&	0.60	$\pm$	0.12	&	2.28	$\pm$	0.46	&	-8.36	$\pm$	0.09	&	0.29	$\pm$	0.06	&	1	\\
5434	&	2	&	149.72563170	&	1.81083000	&	0.752	&	1.20	&	243.50	&	0.44	&	243.93	&	10.70	&	0.227	&	0.85	$\pm$	0.15	&	3.40	$\pm$	0.58	&	-8.14	$\pm$	0.07	&	0.40	$\pm$	0.08	&	1	\\
\hline												
\hline
\multicolumn{16}{c}{Bin 2: Intermediate $\Sigma $ (1.2 $-$ 2.6 Mpc$^{-2})$ $-$ 25 galaxies}\\			
\hline
40594	&	1	&	150.15902710	&	1.97935998	&	0.736	&	1.25	&	18.28	&	0.00	&	18.28	&	10.45	&	0.166	&	0.35	$\pm$	0.10	&	1.34	$\pm$	0.38	&	-8.87	$\pm$	0.12	&	0.32	$\pm$	0.10	&	2	\\
54204	&	2	&	150.37217710	&	2.54916000	&	0.750	&	1.37	&	21.65	&	0.00	&	21.65	&	9.69	&	0.221	&	$<$	0.47	&	$<$	1.88	&	$<$	-8.94	&	$<$	0.79	&	3	\\				
21546	&	2	&	150.05485540	&	2.56947994	&	0.755	&	1.38	&	252.45	&	1.05	&	253.49	&	10.45	&	0.210	&	0.79	$\pm$	0.15	&	3.16	$\pm$	0.59	&	-8.10	$\pm$	0.08	&	0.53	$\pm$	0.11	&	1	\\
29504	&	1	&	150.10209660	&	2.42598009	&	0.727	&	1.40	&	27.83	&	2.61	&	30.44	&	10.44	&	0.179	&	0.56	$\pm$	0.11	&	2.11	$\pm$	0.42	&	-8.84	$\pm$	0.09	&	0.43	$\pm$	0.10	&	2	\\
52621	&	2	&	150.34239200	&	2.57600999	&	0.749	&	1.41	&	52.33	&	2.00	&	54.33	&	10.25	&	0.220	&	0.60	$\pm$	0.15	&	2.38	$\pm$	0.59	&	-8.64	$\pm$	0.11	&	0.57	$\pm$	0.16	&	2	\\
18802	&	1	&	150.00399780	&	2.34864998	&	0.729	&	1.47	&	26.45	&	0.41	&	26.87	&	10.53	&	0.165	&	0.40	$\pm$	0.12	&	1.51	$\pm$	0.44	&	-8.75	$\pm$	0.13	&	0.31	$\pm$	0.10	&	1	\\
30703	&	1	&	150.10815430	&	2.54707003	&	0.726	&	1.53	&	25.48	&	0.70	&	26.18	&	10.80	&	0.168	&	0.33	$\pm$	0.13	&	1.25	$\pm$	0.49	&	-8.68	$\pm$	0.17	&	0.17	$\pm$	0.07	&	1	\\
52339	&	1	&	150.33618160	&	2.35308003	&	0.725	&	1.54	&	36.36	&	0.25	&	36.61	&	10.44	&	0.163	&	0.33	$\pm$	0.10	&	1.24	$\pm$	0.39	&	-8.53	$\pm$	0.14	&	0.31	$\pm$	0.10	&	1	\\
11447	&	1	&	149.87681580	&	2.47651005	&	0.737	&	1.59	&	35.17	&	0.41	&	35.58	&	10.80	&	0.163	&	0.30	$\pm$	0.11	&	1.16	$\pm$	0.41	&	-8.51	$\pm$	0.15	&	0.16	$\pm$	0.06	&	1	\\
11436	&	1	&	149.87672420	&	2.46471000	&	0.731	&	1.62	&	19.14	&	0.53	&	19.67	&	10.54	&	0.178	&	0.30	$\pm$	0.13	&	1.16	$\pm$	0.48	&	-8.77	$\pm$	0.18	&	0.25	$\pm$	0.11	&	1	\\
15692	&	1	&	149.95265200	&	2.51623011	&	0.730	&	1.66	&	51.56	&	1.38	&	52.94	&	10.85	&	0.165	&	0.33	$\pm$	0.12	&	1.27	$\pm$	0.44	&	-8.38	$\pm$	0.15	&	0.15	$\pm$	0.05	&	1	\\
8831	&	2	&	149.82049560	&	1.81175995	&	0.749	&	1.69	&	27.82	&	6.32	&	34.14	&	10.76	&	0.215	&	0.56	$\pm$	0.14	&	2.22	$\pm$	0.58	&	-8.81	$\pm$	0.11	&	0.28	$\pm$	0.08	&	2	\\
12693	&	1	&	149.89828490	&	2.42258000	&	0.733	&	1.74	&	21.60	&	3.18	&	24.78	&	10.06	&	0.166	&	$<$	0.36	&	$<$	1.38	&	$<$	-8.74	&	$<$	0.54	&	3	\\				
26755	&	1	&	150.08918760	&	2.06529999	&	0.725	&	1.78	&	31.71	&	4.10	&	35.82	&	10.89	&	0.168	&	$<$	0.42	&	$<$	1.57	&	$<$	-8.64	&	$<$	0.17	&	3	\\				
16770	&	1	&	149.97004700	&	2.46136999	&	0.732	&	1.79	&	35.84	&	4.73	&	40.58	&	11.05	&	0.166	&	0.31	$\pm$	0.12	&	1.19	$\pm$	0.44	&	-8.47	$\pm$	0.16	&	0.09	$\pm$	0.04	&	1	\\
19191	&	1	&	150.01077270	&	2.56886005	&	0.737	&	2.00	&	20.63	&	3.55	&	24.18	&	10.65	&	0.168	&	0.44	$\pm$	0.11	&	1.69	$\pm$	0.41	&	-8.84	$\pm$	0.10	&	0.27	$\pm$	0.07	&	2	\\
12989	&	1	&	149.90365600	&	2.54447007	&	0.728	&	2.02	&	21.70	&	0.33	&	22.03	&	10.54	&	0.169	&	0.68	$\pm$	0.10	&	2.58	$\pm$	0.38	&	-9.07	$\pm$	0.06	&	0.43	$\pm$	0.07	&	2	\\
8269	&	2	&	149.80523680	&	1.76399004	&	0.744	&	2.05	&	24.73	&	2.55	&	27.27	&	10.08	&	0.208	&	0.62	$\pm$	0.15	&	2.45	$\pm$	0.58	&	-8.95	$\pm$	0.10	&	0.67	$\pm$	0.19	&	2	\\
43989	&	1	&	150.18231200	&	1.70081997	&	0.735	&	2.16	&	109.76	&	0.83	&	110.59	&	10.64	&	0.181	&	0.54	$\pm$	0.12	&	2.05	$\pm$	0.45	&	-8.27	$\pm$	0.10	&	0.32	$\pm$	0.08	&	1	\\
46397	&	2	&	150.21011350	&	2.31167006	&	0.748	&	2.16	&	133.30	&	1.74	&	135.03	&	10.80	&	0.216	&	1.19	$\pm$	0.17	&	4.76	$\pm$	0.68	&	-8.55	$\pm$	0.06	&	0.43	$\pm$	0.07	&	1	\\
14696	&	1	&	149.93466190	&	2.55611992	&	0.729	&	2.16	&	24.01	&	3.96	&	27.96	&	11.06	&	0.172	&	0.39	$\pm$	0.11	&	1.48	$\pm$	0.42	&	-8.72	$\pm$	0.12	&	0.11	$\pm$	0.03	&	2	\\
63717	&	1	&	150.64860540	&	1.90240002	&	0.737	&	2.47	&	38.75	&	3.10	&	41.84	&	10.13	&	0.176	&	0.36	$\pm$	0.11	&	1.38	$\pm$	0.42	&	-8.52	$\pm$	0.13	&	0.51	$\pm$	0.18	&	2	\\
20173	&	1	&	150.02958680	&	2.30753994	&	0.726	&	2.48	&	43.81	&	0.61	&	44.42	&	10.90	&	0.167	&	0.36	$\pm$	0.13	&	1.36	$\pm$	0.50	&	-8.49	$\pm$	0.16	&	0.15	$\pm$	0.06	&	1	\\
9757	&	1	&	149.84196470	&	2.50101995	&	0.734	&	2.58	&	32.00	&	1.44	&	33.44	&	9.78	&	0.165	&	$<$	0.32	&	$<$	1.23	&	$<$	-8.56	&	$<$	0.67	&	3	\\				
13409	&	2	&	149.91070560	&	2.55461001	&	0.753	&	2.60	&	51.37	&	7.00	&	58.37	&	10.78	&	0.223	&	0.62	$\pm$	0.15	&	2.48	$\pm$	0.59	&	-8.63	$\pm$	0.10	&	0.29	$\pm$	0.08	&	2	\\
\hline	
\hline
\multicolumn{16}{c}{Bin 3: High $\Sigma $ ($>$ 2.6 Mpc$^{-2})$ $-$ 15 galaxies}\\
\hline
9173	&	1	&	149.82911680	&	2.41458011	&	0.730	&	3.00	&	26.55	&	0.30	&	26.85	&	10.43	&	0.170	&	0.47	$\pm$	0.10	&	1.77	$\pm$	0.39	&	-8.82	$\pm$	0.10	&	0.40	$\pm$	0.12	&	2	\\
17344	&	1	&	149.97996520	&	1.76528001	&	0.735	&	3.27	&	32.76	&	1.06	&	33.81	&	10.44	&	0.180	&	0.32	$\pm$	0.13	&	1.23	$\pm$	0.48	&	-8.56	$\pm$	0.17	&	0.31	$\pm$	0.13	&	1	\\
11235	&	1	&	149.87269590	&	2.49549007	&	0.732	&	3.31	&	18.41	&	0.43	&	18.84	&	10.33	&	0.168	&	0.33	$\pm$	0.10	&	1.27	$\pm$	0.39	&	-8.83	$\pm$	0.14	&	0.37	$\pm$	0.13	&	2	\\
18732	&	1	&	150.00265500	&	2.68545008	&	0.735	&	3.41	&	44.57	&	1.49	&	46.06	&	10.65	&	0.180	&	$<$	0.43	&	$<$	1.64	&	$<$	-8.55	&	$<$	0.27	&	3	\\				
10717	&	1	&	149.86195370	&	2.47996998	&	0.727	&	3.56	&	26.65	&	0.23	&	26.88	&	10.00	&	0.174	&	$<$	0.41	&	$<$	1.54	&	$<$	-8.76	&	$<$	0.61	&	3	\\				
13621	&	1	&	149.91462710	&	2.54237008	&	0.729	&	3.93	&	24.74	&	0.43	&	25.18	&	10.71	&	0.173	&	0.39	$\pm$	0.10	&	1.48	$\pm$	0.39	&	-8.77	$\pm$	0.12	&	0.22	$\pm$	0.06	&	2	\\
28053	&	1	&	150.09541320	&	2.25518990	&	0.727	&	4.28	&	28.51	&	1.08	&	29.59	&	11.00	&	0.167	&	0.48	$\pm$	0.10	&	1.82	$\pm$	0.39	&	-8.79	$\pm$	0.09	&	0.15	$\pm$	0.03	&	2	\\
20897	&	1	&	150.04356380	&	2.32572007	&	0.727	&	4.59	&	18.52	&	0.14	&	18.66	&	10.51	&	0.165	&	0.40	$\pm$	0.10	&	1.52	$\pm$	0.38	&	-8.91	$\pm$	0.11	&	0.32	$\pm$	0.09	&	2	\\
13512	&	1	&	149.91262820	&	2.52716994	&	0.737	&	4.69	&	27.02	&	1.70	&	28.73	&	9.48	&	0.177	&	$<$	0.33	&	$<$	1.26	&	$<$	-8.64	&	$<$	0.81	&	3	\\				
23357	&	1	&	150.06848150	&	2.32342005	&	0.729	&	4.81	&	33.68	&	0.00	&	33.68	&	10.40	&	0.159	&	0.31	$\pm$	0.11	&	1.17	$\pm$	0.41	&	-8.54	$\pm$	0.15	&	0.32	$\pm$	0.12	&	1	\\
25873	&	2	&	150.08401490	&	2.64053988	&	0.747	&	5.11	&	26.77	&	1.72	&	28.48	&	10.32	&	0.216	&	$<$	0.45	&	$<$	1.79	&	$<$	-8.80	&	$<$	0.46	&	3	\\				
32530	&	2	&	150.11770630	&	2.26676011	&	0.751	&	5.31	&	28.77	&	2.30	&	31.07	&	11.15	&	0.152	&	0.41	$\pm$	0.11	&	1.65	$\pm$	0.43	&	-8.73	$\pm$	0.11	&	0.10	$\pm$	0.03	&	2	\\
10837	&	1	&	149.86451720	&	2.48399997	&	0.729	&	5.37	&	17.12	&	0.00	&	17.12	&	9.80	&	0.168	&	0.49	$\pm$	0.10	&	1.86	$\pm$	0.39	&	-9.04	$\pm$	0.09	&	0.75	$\pm$	0.20	&	2	\\
32328	&	2	&	150.11659240	&	2.26526999	&	0.749	&	5.52	&	20.10	&	0.38	&	20.48	&	10.20	&	0.152	&	0.41	$\pm$	0.10	&	1.65	$\pm$	0.42	&	-8.91	$\pm$	0.11	&	0.51	$\pm$	0.15	&	2	\\
17691	&	1	&	149.98559570	&	2.55435991	&	0.728	&	10.48	&	35.76	&	0.82	&	36.58	&	10.80	&	0.165	&	0.59	$\pm$	0.10	&	2.22	$\pm$	0.38	&	-8.78	$\pm$	0.07	&	0.26	$\pm$	0.05	&	2	\\
\enddata
\tablenotetext{a}{SG: ALMA Science Goal}
\tablenotetext{b}{$f_\mathrm{gas} = $ M$_{\mathrm{ISM}}$/(M$_{\mathrm{ISM}}$ + M$_*$)}
\tablenotetext{c}{1 - flux found from integrated aperture; 2 - flux found from peak pixel; 3 - 3$\sigma$ upper limit}
\end{deluxetable*}
\end{longrotatetable}

\bibliography{reference}

\end{document}